\documentclass[useAMS,usenatbib]{mn2e}

\usepackage{epsfig}
\usepackage{amssymb}
\usepackage{natbib,multirow}

\def\deg{\ifmmode^\circ\else$^\circ$\fi}
\def\Msun{\ifmmode{\mathrm M_\odot}\else{M$_\odot$}\fi}

\def\ah{\ifmmode{^\textrm{\scriptsize h}}\else{$^\textrm{\scriptsize h}$}\fi}
\def\am{\ifmmode{^\textrm{\scriptsize m}}\else{$^\textrm{\scriptsize m}$}\fi}
\def\as{\ifmmode{^\textrm{\scriptsize s}}\else{$^\textrm{\scriptsize s}$}\fi}

\title[The trace of a substantial assembly of massive E--S0's in galaxy number counts]{The trace
 of a substantial assembly of massive E--S0 galaxies at $0.8<z<1.5$ in galaxy number counts}

\author[Prieto \& Eliche-Moral]
{\parbox{\textwidth}{Mercedes Prieto$^{1,2}$\thanks{E-mail: mpm@iac.es} and M.\,Carmen Eliche-Moral$^{3}$}\\ \\
$^{1}$Instituto de Astrof\'{\i}sica de Canarias, C/ V\'{\i}a L\'actea, E-38200 La Laguna, Tenerife, Spain\\
$^{2}$Departamento de Astrof\'{\i}sica, Universidad de La Laguna, Avda.\ Astrof\'{\i}sico Fco.\,S\'anchez, E-38200 La Laguna, Tenerife, Spain\\
$^{3}$Departamento de Astrof\'{\i}sica y CC.\,de la Atm\'{o}sfera, Universidad Complutense de Madrid, E-28040 Madrid, Spain}

\begin{document}

\date{Accepted 2015 May 04. Received 2015 May 04; in original form 2014 August 05}
\pagerange{\pageref{firstpage}--\pageref{lastpage}} \pubyear{2000}

\maketitle

\label{firstpage}


\begin{abstract}
$K$-band galaxy number counts (GNCs) exhibit a slope change at $K\sim 17.5$\,mag not present in optical bands. To unveil the nature of this feature, we have derived the contribution of different galaxy types to the total $K$-band GNCs at $0.3<z<1.5$ by redshift bins and compared the results with expectations from several galaxy evolutionary models. We show that the slope change is caused by a sudden swap of the galaxy population that numerically dominates the total GNCs (from quiescent E--S0's at $K< 17.5$\,mag to blue star-forming discs at fainter magnitudes), and that it is associated with a flattening of the contribution 
of the E--S0's at $0.6<z<1$ to the total GNCs. We confirm previous studies showing that models in which the bulk of massive E--S0's have evolved passively since $z>2$ cannot predict the slope change, whereas those imposing a relatively late assembly on them ($z<1.5$) can reproduce it. The $K$-band GNCs by redshift bins and morphological types point to a progressively definitive build-up of $\sim$50\% of this galaxy population at $0.8<z<1.5$, which can be explained only through the major mergers reported by observations. We conclude that the slope change in total $K$-band GNCs is a vestige of the definitive assembly of a substantial fraction of present-day massive E--S0's at $0.8<z<1.5$.
\end{abstract}


\section{Introduction}
\label{sec:introduction}

Galaxy number counts (GNCs, i.e.\ the number of galaxies per apparent magnitude and area unit on the sky) were originally conceived to constrain cosmological models, a purpose for which they are still being used \citep[see][]{1935ApJ....82..302H,2008Ap&SS.318...13H,2010JCAP...01..004E,2015arXiv150304033S}. However, it was soon realized that they are more sensitive to the processes driving galaxy evolution than to the assumed cosmological model \citep{1974ApJ...194..555B,2009ApJ...705.1462H}. 

Although the estimates of luminosity functions (LFs) have the advantage over GNCs of showing the intrinsic evolution of each galaxy population directly, LF estimates are still tricky owing to the assumptions adopted for the statistical method, the cosmological model, and the \emph{e}-correction \citep{1997AJ....114..898W}, as well as the observational selection effects that can make any galaxy sample incomplete \citep{2011A&ARv..19...41J}. Completeness in spectroscopic surveys at intermediate-to-high redshifts is still challenging, so most studies use photometric redshifts to obtain large complete samples \citep[e.g.\ ][]{2005ApJ...630...82P,2011ApJ...742...24L,2015arXiv150206602S}. But small errors in the photometric redshifts translate into large uncertainties in the rest-frame fluxes, affecting more noticeably the LF estimates when the object statistics are low. Consequently, the LFs of bright (i.e.\ massive) galaxies tend to be poorly estimated and their evolution is less constrained 
\citep[][]{2010A&A...514A...6G,2015arXiv150300004K}. Additionally, there are very few studies deriving $k$-corrections from real data, and they are usually limited to low-to-intermediate redshifts, certain bands, and/or some specific galaxy types, so LFs are affected by these uncertainties too \citep[e.g.\ ][]{2009MNRAS.398.1549R,
2009A&A...497..743H,2010MNRAS.405.1409C}. These uncertainties in the $k$-corrections also increase the difficulty of modelling. In fact, current models can predict the galaxy luminosity functions at $z\sim 0$, but not their precise evolution with redshift \citep{2011MNRAS.413..101G}.

On the contrary, GNCs are free from these uncertainties and have higher statistical significance, so they are quite robust observational diagnostics for discriminating between different evolutionary models.Nevertheless, despite being apparently simple, GNCs summarize the effects of different galaxy evolutionary processes taking place at different redshifts, so they still represent a challenge for current galaxy evolution models, which have difficulty in reproducing total GNCs from NUV to NIR simultaneously
 \citep{2006MNRAS.366..858K,2007MNRAS.376....2K,2009A&A...505.1041G}. This difficulty arises because the slope of
 GNCs in optical bands is almost featureless and nearly Euclidean for a wide magnitude range \citep[$d\log(N)/dm\sim 0.6$, see][]{1971phco.book.....P,2001AJ....122.1104Y}, but it exhibits a noticeable slope change in NIR GNCs at $K\sim 17.5$ mag \citep[see][]{1993ApJ...415L...9G,1998ApJ...505...50B,
2000MNRAS.311..707M,2003ApJ...595...71C,2009ApJ...696.1554C,2007AJ....133.2418I,2009A&A...494...63B}. This slope change is also reported in the $J$ and $H$ bands, although it is less noticeable \citep{2009ApJ...696.1554C}. 

Many studies have tried to derive the processes responsible for this feature through modelling, but with little success. In general, whereas UV and blue optical GNCs require strong luminosity evolution of galaxy populations and/or high merger rates to be fitted 
\citep{1997MNRAS.288..404H,2000ApJ...542L..79G,2001A&A...367..788F,2002ApJ...578..675N,2009ApJ...705.1462H}, red optical and NIR GNCs basically follow a pure luminosity evolution (PLE) trend for all galaxy types  
\citep[at least until the magnitudes of the slope change, see][]{1996MNRAS.281..953P,1998MNRAS.298..483H,2000ApJ...540...81T,
2001AJ....121..598M,2001A&A...368..787H,2001AJ....121..598M,2001ApJ...559..592T,2004ApJ...600L.135S}. No consistent 
evolutionary scenario has so far been proposed to reproduce GNCs simultaneously by redshift bin and morphological type 
in all bands.

\citet{1998ApJ...505...50B} compared deep GNCs in the $J$ and $K$ bands with the expectations of different galaxy evolutionary models. These authors had already attributed the slope change in the NIR bands to the evolution experienced by the ancestors of present-day bright quiescent galaxies, because they had found a deficit of galaxies with $J-K$ colours typical of bright passively-evolving objects at $1<z<3$ with respect to non-evolutionary models that indicated that these galaxies were much bluer at these epochs than at $z<1$ \citep[see also][]{1999Ap&SS.267..343G}. It was \citet[CH03 hereafter]{2003ApJ...595...71C} who first suggested that the slope change could be reproduced by delaying the formation of massive early-type galaxies (E--S0's with masses $M_*> 10^{11}\Msun$) down to $z\lesssim 2$. However, their model failed to reproduce GNCs in blue optical bands unless a transient population of star-forming galaxies was included ad hoc at all redshifts. In any case, the assembly epoch proposed for massive early-type galaxies was strikingly late compared to paradigms (mainly monolithic collapse at $z>3$) that were prevalent in those years.

\citet{2006MNRAS.366..858K} showed that PLE models for early-type galaxies overpredict the population of massive E--S0's at high redshift. They supported the conclusions by previous authors that, in order to reproduce GNCs data, either the majority of massive E--S0's should be assembled at late epochs, or the ancestors of these galaxies should have suffered strong dust extinction phases at $z\gtrsim 1$. A coeval study proved that both mechanisms were indeed required to reproduce total GNCs simultaneously in the $U$, $B$, and $K$ bands \citep[][EM06 henceforth]{2006ApJ...639..644E}. The appearance of the bulk of massive E--S0's on the cosmic scenario at a quite late redshift ($z=1.5$) was necessary to reproduce the slope change in the NIR GNCs, but a moderate dust extinction in the early evolutionary stages of these galaxies was necessary to remove `bump' signatures in the GNCs of blue bands caused by the young stellar populations contained in these galaxies at $z\sim 1.5$. The EM06 model did not identify the mechanism responsible for the E--S0 assembly, but these authors suggested that the high extinction required in the early phases of this galaxy population may indicate gas-rich major mergers.

Later studies based on previous models estimated that the present-day number density of massive early-type galaxies should decrease by a factor of $\sim$2--3 at $z\sim 1$ to reproduce multiwavelength GNC data \citep{2009ApJ...696.1554C}. This association of the slope change in NIR GNCs and the late build-up of massive red objects were supported by \citet[][]{2009A&A...494...63B}, who showed via GNCs by redshift bins, that the slope change in NIR GNCs was caused by a prominent decrease in the number density of $L\sim L^*$ objects at $0.8<z<1$. However, neither the galaxy population nor the mechanism responsible for the slope change in the $K$-band GNCs have been observationally identified or confirmed yet. 

In order to test observationally  whether the slope change in NIR GNCs is really a trace of the recent build-up of massive E--S0's or not, we have analysed the contribution  to the total $K$-band GNCs of the various galaxy morphological 
types (in particular, the E--S0's) at $0.3<z<1.5$.  The results obtained have been compared with the predictions of several galaxy evolution models in order to shed some light on the mechanisms responsible for this feature in NIR GNCs. \citet{2009A&A...505.1041G} proposed that the discrepancies between models and GNC data could be evaluated in more detail by dividing the counts into redshift bins, as was done by \citet{2009A&A...494...63B}. GNCs in redshift bins can disentangle global evolutionary effects that may depend on the redshift, and that are mixed in the total GNC, thereby at the same time avoiding the large uncertainties associated with LF estimates (see above). We have therefore also derived the GNCs of E--S0's by wide redshift bins in our sample in order to compare with the expectations from the models. 

The paper is organized as follows. In \S\ref{sec:data} we describe the data and the procedure used to derive GNCs by redshift bin and by morphology in the $K$ band. In \S\ref{sec:model}, we provide a brief outline of the GNC models with which we have compared our observational results. The results are described and discussed in \S\ref{sec:results}. Section \ref{sec:GNCsobserved} discusses the results derived directly from the observational GNCs obtained here concerning the nature of the slope change in NIR GNCs. In \S\ref{sec:gncspredicted}, we show the clues to the evolutionary mechanisms of massive E--S0's that arise from the comparison of GNC data with the predictions of the models. The conclusions are finally provided in \S\ref{sec:conclusions}. We use a concordant cosmology throughout the paper \citep[$\Omega_M = 0.3$, $\Omega_\Lambda = 0.7$, $H_0 = 70$ km s$^{-1}$ Mpc$^{-1}$, see][]{2007ApJS..170..377S}. All magnitudes have been converted into the Vega system.

\begin{figure*}
\begin{center}
\includegraphics[width=\textwidth,angle=0, bb = 10 5 481 198, clip]{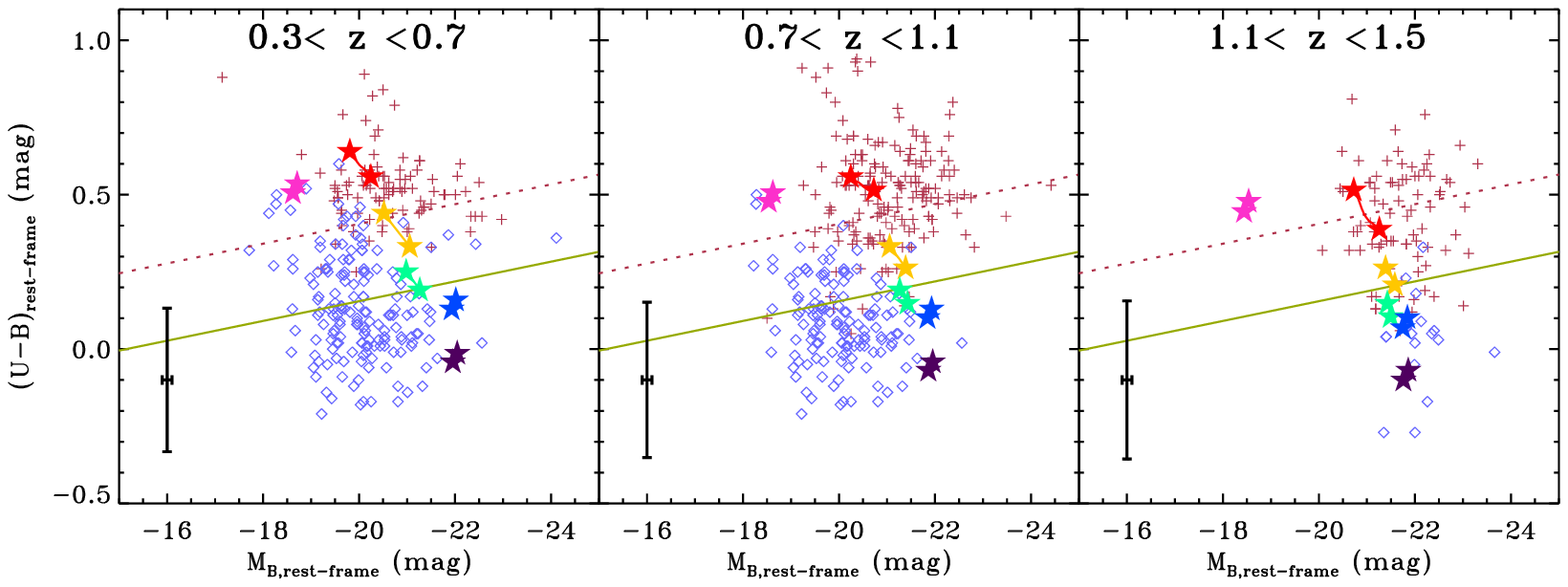}
\includegraphics[width=\textwidth,angle=0, bb = -5 0 481 160, clip]{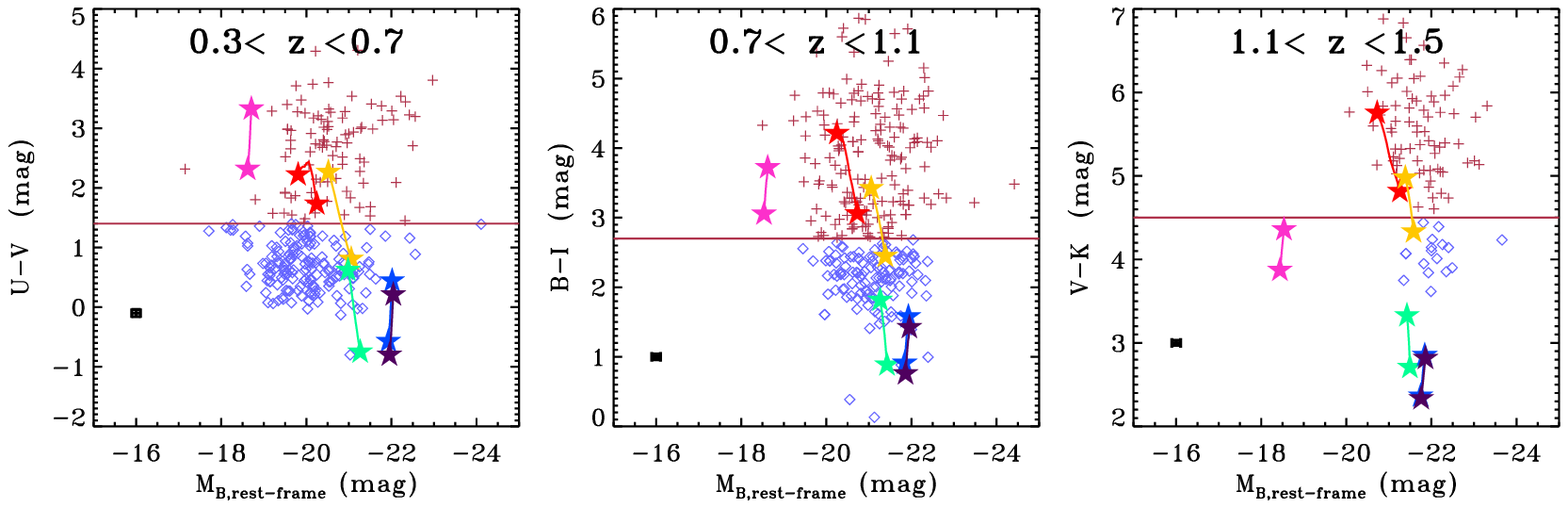}
\caption{ Colour--magnitude diagrams for the P13 observational sample with $K< 19.5$\,mag in the three wide redshift bins  ($0.3<z<0.7$, $0.7<z<1.1$, and $1.1<z<1.5$) considered in that study. Red and blue galaxies, according to the criteria used by P13, are plotted with \emph{red crosses} and \emph{blue diamonds}, respectively. The median errors for the observational data in each axis are plotted in the bottom-left corner of each panel. \textbf{Top panels}: rest-frame ($U-B$) colour vs.\ rest-frame absolute magnitude in the $B$ band, $M_{B\mathrm{,rest-frame}}$, for the galaxies in the P13 sample, using the values available in the Rainbow database. The \emph{red dotted line} corresponds to the fit to the red sequence E--S0 galaxies obtained by \citet{2000ApJ...541...95V}, and the \emph{green solid line} is the same line shifted downwards by 0.25 mag to pass through the 
valley between the  red and blue galaxies \citep[usually used as a cut to distinguish between red and blue populations; e.g.\ ][]{2006ApJ...647..853W}. \textbf{Bottom panels}: available apparent colours nearest to the rest-frame ($U - B$) colour at each redshift interval vs.\,$M_{B\mathrm{, rest-frame}}$ for the same galaxies. \emph{Horizontal red solid lines}: colour cuts defined by P13 to distinguish between red and blue galaxies on the basis of apparent colours. \textbf{Lines limited by stars}: 
theoretical trends followed by different galaxy types within the redshift interval of each frame, modelled assuming standard SFHs starting at $z_f=3$ and characteristic physical properties according to observations.  These trends have been obtained with the stellar population synthesis models by \citet[see the text for details]{2003MNRAS.344.1000B}. A stellar mass of $\log(M_* / \Msun) = 11$ has been assumed for all models. \emph{Red}: E--S0 galaxy. \emph{Yellow}: Sa-Sab. \emph{Green}: Sb-Sbc. \emph{Blue}: Sc-Scd. \emph{Purple}: Sd-Irr. \emph{Pink}: Dust-reddened starburst galaxy with $A_V=3.0$\,mag. [\emph{A colour version is available in the online version.}]
\/} \label{fig:cmd}
\end{center}
\end{figure*}

\section{Data and GNC estimation}
\label{sec:data}

This study is based on the data, selection, and classification procedures by \citet[P13 hereafter]{2013MNRAS.428..999P}. 
Detailed information is provided in the original paper, so here we provide only a brief description. 

We have used broad band photometric data ($U$, $B$, $V$, $I$, $J$, and $K$) from the Rainbow Extragalactic Database\footnote{Rainbow Extragalactic Database:\\https://rainbowx.fis.ucm.es/\-Rainbow\_Database/\-Home.html} 
\citep{2011ApJS..193...13B} over an area of $\sim$$155$\,arcmin$^{2}$ in the Groth Strip \citep[$\alpha=14^{\rm h} 16^{\rm m} 38.8^{\rm s}$ and $\delta=52^\circ 16' 52''$, see][]{1994AAS...185.5309G,1999AJ....118...86R,2002ApJS..142....1S}. The database contains photometric data from the UV to the FIR over this field, as well as diagnostics derived from the fit of the spectral energy distributions of nearly 80\,000 galaxies selected in IRAC bands. We have used photometric redshifts available in this database, with an accuracy of $\langle\Delta z/(1+z)\rangle \sim 0.03$ (see fig.\,2 in P13). The precision of the 
photometric redshifts (measured as the 68\% confidence interval in the probability distribution function of the redshift estimates around the most probable photometric redshift computed with the Rainbow code) differ according to redshift, but are typically below $\Delta z_\mathrm{phot}\sim 0.08$ \citep[see table\,2 in][]{2011ApJS..193...30B}. The depths reached are typically $B\sim 26$ and $K\sim 21$\,mag. We started with a sample of $\sim$850 sources with $K<19.5$ mag at 90\% completeness in terms of detection. This sample is complete in detection at the 100\% level for $K<18.5$ mag, thus ensuring total completeness of the GNCs down to this limiting magnitude. Consequently, the data adequately cover the range of magnitudes in which the slope change in the $K$-band GNCs appears without requiring efficiency corrections. We preferred to avoid  such corrections because they often fail to recover the true GNC distribution owing to data errors and unrealistic assumptions adopted in the models developed to estimate them \citep[see comments concerning this in, e.g.,][]{2003ApJ...595...71C,2006ApJ...639..644E}.

In order to understand the differences between the GNCs in NIR and UV--blue bands, we first classified the galaxies according to their colours into those belonging to the red sequence or the blue cloud. This bimodality in the basic properties of galaxies can be traced up to $z\sim 2$ \citep{2007MNRAS.377.1717K,2009ApJ...691.1879W,2011ApJ...739...24B,2011ApJ...735...86W}. The rest-frame $U-B$ colour is usually used for this purpose. This colour traces the 4000\,\AA\ break, a spectral feature characteristic of evolved stellar populations, so it is a good discriminator between really quiescent (red) and star-forming (blue) galaxies. However, we had to perform this classification in terms of apparent, instead of rest-frame, colours because of the large uncertainties in these latter that result from the propagation of the errors in the photometric redshifts and magnitudes. 

We illustrate this problem in Figure \ref{fig:cmd}. In the top panels, we plot the distribution of our galaxies in the rest-frame $U-B$ vs.\ $M_B$ diagram for three wide redshift bins ($0.3<z<0.7$, $0.7<z<1.1$, and $1.1<z<1.5$). We have overplotted the linear fit performed to describe the location of the red sequence by \citet{2000ApJ...541...95V}, as well as the limit used by \citet{2006ApJ...647..853W} to isolate the galaxies on the red sequence from those lying in the blue cloud. We have represented the median error bars in both axes in each panel. Attending to the typical errors in our rest-frame $U-B$ estimates, it is obvious that we cannot robustly assign a colour class to the galaxies on the basis of a colour--magnitude cut in this diagram 
as in the one defined by \citeauthor{2006ApJ...647..853W}. This prevented us from trusting these rest-frame colours for making the red--blue distinction in our sample. 

We used instead the observed apparent colour from our dataset that traces the rest-frame $U-B$ more closely at the centre of the three wide redshift bins under consideration ($U-V$, $B-I$, and $V-K$ for the low-, middle-, and high-redshift intervals, respectively). We show the distribution of these colours for our galaxies  as a function of the rest-frame $M_B$ in the bottom panels of Fig.\,\ref{fig:cmd} for the three redshift bins. The apparent colour distributions also exhibit the well-known bimodal distribution of galaxies into red and blue populations (as do the rest-frame $U-B$ diagrams), with the advantage of having much lower errors in the colours (compare the median error bars in these diagrams with those in the corresponding rest-frame diagrams at the top panels). Therefore, the classification into red and blue galaxies is much more reliable for our data if it is defined according to these apparent colours. 

We thus defined the colour cuts assuming Gaussian-shaped distributions for the red and blue galaxy populations in the histogram of the apparent colour corresponding to each redshift bin. The cut was defined as the  point where the fitted Gaussian distributions crossed. The blue and red distributions were clearly defined  in the first two redshift bins, but not in the last one. We defined the colour cut in the highest redshift interval by ensuring that it selected similar galaxy types to those defined in the lower redshift bins. For this purpose, we used the evolutionary tracks followed in colour--redshift diagrams by characteristic galaxy types, as derived from the stellar population synthesis models by \citet{2003MNRAS.344.1000B}. The resulting cuts were $(U-B)=1.4$\,mag for $0.3<z<0.7$, $(B-I)=2.7$\,mag for $0.7<z<1.1$, and $(V-K)=4.5$\,mag (see Fig.\,3 in P13).

In Fig.\,\ref{fig:cmd} we have also plotted the locations expected for some of these models in the 
colour--magnitude diagrams for each redshift interval. We overplot the colours and magnitudes 
expected for a typical E--S0, Sa--Sab, Sb--Sbc, Sc--Scd, Sd--Irr, and dust-reddened starburst galaxies, as defined by the parameterizations provided for the star formation histories (SFHs) and the characteristic 
physical properties (metallicity, dust extinction, e-folding timescale for the star formation) of each type 
defined in \citet[EM10 henceforth]{2010A&A...519A..55E}, based on observational studies (see \S\S\ref{sec:PLE}
 and \ref{sec:EM10} for more information). The star formation is set to start at $z_f=3$ for all types, and a 
stellar mass of $10^{11}\Msun$ is assumed for them all. The top panels of the figure shows that E--S0's and
 dust-reddened star-forming galaxies are expected to lie in the region of `red galaxies' according to the
 colour cuts defined in the rest-frame $U-B$ diagrams for all redshifts, whereas early-type spirals would lie
 in the green valley and late-type spirals in the blue cloud. According to the cut defined by 
\citet{2006ApJ...647..853W}, E--S0's, dust-reddened starbursts, and Sa--Sab's would be identified as 
`red galaxies', and spirals of types later than Sb lie would be `blue galaxies'. In the bottom 
panels, the equivalent colour cuts defined in P13 in the apparent colours basically distinguish between the 
same galaxy models, thus ensuring that these colour cuts select homogeneous samples of galaxy populations in
 the whole redshift interval in a similar way to those defined in terms of rest-frame colours (see 
Figs.\,3 and 4 and \S\,3.1 in P13 for more details).

Additionally, colour--colour diagrams were used to remove stars from the sample. We have finally got a 
red galaxy sample with 302 systems and a blue one of 536 objects with $K<19.5$\,mag.  The blue galaxy 
sample basically consists of star-forming spirals. The P13 study was centered on the build-up of the red sequence, 
so no morphological classification was performed for blue galaxies. In the present study, we show that the 
slope change in the $K$ band is not related to this galaxy population, in the sense that the GNCs of blue galaxies 
do not exhibit any change of trend at $16<K<19.5$ (see \S\ref{sec:GNCsobserved}). Therefore, we will consider that
 blue galaxies make up an independent galaxy type throughout the paper and we will not distinguish different 
morphological types among them.

In the red sample, however, we have used the visual morphological classification developed by P13. The 
classification criteria were centered on the global morphology of the galaxy (if it was a spheroid, or a 
disc-dominated object) and the distortion level of the whole galaxy body (whether the galaxy was `regular',
 in the sense of poorly distorted morphology,  or `irregular', meaning that it exhibited a significant distortion). 
The definitive visual classification accounted for the median results of three independent classifiers. We used 
surface brightness isophotes and surface maps in the $I$ band (besides the images) to identify structural 
distortions and morphological features. This band was selected for the visual classification because it is the 
reddest optical band with the best spatial resolution available in the data set.

Spheroid-dominated galaxies were defined as those exhibiting central light concentrations resembling those 
of nearby E--S0's  (i.e.\ the central bulge controlled the surface brightness profile up to at least one third of the galaxy size). We identified as `irregular or distorted' galaxies only those with morphological 
features characteristic of major mergers, such as train-wreck morphologies, equal-length tidal tails, or 
double nuclei objects with high and similar brightness, following the methodology by \citet[]{2009ApJ...697.1971J}. 
We thus avoided the inclusion of galaxies at an early stage of a major encounter, undergoing minor mergers, or
 experiencing inhomogeneous starbursts along their discs into this `irregular' class. However, some
 contamination by these objects cannot be excluded. Consequently, attending to the global morphology and 
structural distortion level, P13 defined six major exclusive classes of red galaxies:

\begin{itemize}
 
\item Compact galaxies.-- Galaxies exhibiting compact morphologies, according to the seeing of the images. 
The number of these objects in the sample was negligible.

\item Regular Spheroids.-- Galaxies dominated by a central spheroid, with regular isophotes.
 
\item Regular Discs.-- Galaxies dominated by a disc component with regular isophotes. 

\item Irregular Spheroids.-- Galaxies with irregular isophotes, dominated by an spheroidal component.

\item Irregular Discs.-- Galaxies with irregular isophotes, dominated by a disc.

\item Non-Classified.-- Galaxies that could not be classified owing to their faintness or noise. The number of these objects was negligible too.

\end{itemize}

The visual classification was validated by comparing the results with the automatic classifications resulting from two quantitative parameters known to be proxies of the concentration and distortion level. The tests developed to estimate the robustness of the visual classification are detailed in \S4.2 of P13.

We have estimated the GNCs for red regular spheroid-dominated (basically, E--S0's), and red regular disc-dominated (hereafter, `red disc') galaxies, as well as for red galaxies with highly distorted morphologies and blue ones. The GNCs obtained at $16 < K < 18.5$\,mag are listed in Table\,\ref{Tab:GNCs}, both in total and for each galaxy class. The errors listed in the Table account quadratically for the statistical counting uncertainties at the 84.13\% confidence level \citep{1986ApJ...303..336G} and the classification errors. The uncertainties due to classification are estimated as the square root of the quadratic mean of the statistical errors of the three independent classifications.

In Table\,\ref{Tab:GNCsES0}, we list the $K$-band GNCs that we have obtained for red regular spheroidal-dominated systems (E--S0's) in different redshift bins and for the whole redshift range from $z=0.3$ to $z=1.5$. The errors listed in the table correspond to the quadratic summation of the squared uncertainties associated with statistical counting, classification error, and uncertainties associated with cosmic variance, which have been estimated as described below. We have not accounted for the uncertainties due to redshift errors because they were negligible for the redshift bins under consideration (see P13 for more details). 

We have derived cosmic variance estimates associated with our sample using the model by \citet{2011ApJ...731..113M}, 
which predicts the cosmic variance associated with a galaxy population of a given stellar mass considering a $\Lambda$CDM 
cosmology and the galaxy bias. These authors have kindly made a cosmic variance calculator publicly available
 \citep[{\tt QUICKCV}, see][]{2014ascl.soft02012N}, which provides the root of the cosmic variance of galaxies 
($\sigma_\mathrm{gg}$) expected for different stellar mass bins in different redshift intervals for a survey of a given geometry. 

The objects dominating the GNCs of quiescent galaxies in each redshift bin are those with $L\sim L^\ast$, because 
the LF of these objects  always has a decreasing slope at faint magnitudes at all redshifts up to $z\sim 2$ \citep[in 
fact, $\alpha>-0.6$ at $z\lesssim 2$ for this galaxy population, see][]{2007A&A...476..137A,2007MNRAS.380..585C}. The 
stellar mass typical of quiescent galaxies with $L\sim L^\ast$ rises from $\sim 2\times10^{11}\Msun$ to
 $\sim3$--$5\times10^{11}\Msun$ from $z\sim 0$ to $z\sim 1.5$, considering the typical brightening experienced
 by this galaxy population and the evolution of the mass-to-light ratio in the $K$ band derived for these objects 
by \citet{2007A&A...476..137A}. This means that the objects that control the GNCs of E--S0's at all magnitudes
 ($L\gtrsim L^\ast$) have stellar masses $M_\ast \gtrsim 10^{11}\Msun$ up to $z\sim 1.5$. For galaxies with stellar 
masses $\log(M_\ast / \Msun) > 11$ (henceforth, simply `massive'), the model by
 \citeauthor{2011ApJ...731..113M} predicts that the fractional error ($\sigma_\mathrm{gg}/ N$) is up to $\sim$$28$--30\%
 for the redshift bins and the field considered here. Therefore, our GNCs of E--S0's by redshift bin are affected by 
cosmic variance uncertainties of $\sim$$30$\% at most. We have included this maximum uncertainty in the errors of the
 GNCs of E--S0's by redshift bin in Table\,\ref{Tab:GNCsES0}, as noted above.

Table\,\ref{Tab:GNCsES02} also shows results on the GNCs of E--S0's by redshift bin, but for wider redshift 
intervals ($0.4<z<0.8$, $0.6<z<1.0$, and $0.8<z<1.2$) than those used in Table\,\ref{Tab:GNCsES0}. We have 
derived them in order to reduce the errors associated with the distributions of GNCs of E--S0's in narrow redshift
bins, by raising the statistics and reducing the cosmic variance uncertainties. This allows us to derive more robust
conclusions from the comparison with the models, as we show in \S\ref{sec:gncsES0zbins}. The errors in 
Table\,\ref{Tab:GNCsES02} also account for statistical counting and classification errors, as well as for cosmic variance uncertainties, which are reduced to $\lesssim$15\% for these redshift bins. Note that these wide redshift intervals overlap. 

\begin{table*}
 \begin{minipage}{\textwidth}
\caption{$K$-band GNCs, both total and by galaxy type as defined in P13\label{Tab:GNCs}}
{\scriptsize
  
\begin{center}
  \begin{tabular}{lccccccccccccccccccc}
  \hline
 & \multicolumn{3}{c}{Red normal bulge-dominated$^\mathrm{\star}$} & & \multicolumn{3}{c}{Red highly distorted} & & \multicolumn{3}{c}{Red normal disc-dominated} & & \multicolumn{3}{c}{Blue galaxies} & & \multicolumn{3}{c}{Total}\\\cline{2-4}\cline{6-8}\cline{10-12}\cline{14-16}\cline{18-20}\\\vspace{-0.5cm}\\
\multicolumn{1}{c}{$K^\mathrm{\ast}$} & \multicolumn{1}{c}{$\log N^\mathrm{\dag}$} & \multicolumn{1}{c}{$\Delta_{\mathrm{l}}^\mathrm{\ddag}$} & \multicolumn{1}{c}{$\Delta_{\mathrm{u}}^\mathrm{\ddag}$} & & 
\multicolumn{1}{c}{$\log N$} & \multicolumn{1}{c}{$\Delta_{\mathrm{l}}$} & \multicolumn{1}{c}{$\Delta_{\mathrm{u}}$} &  & 
\multicolumn{1}{c}{$\log N$} & \multicolumn{1}{c}{$\Delta_{\mathrm{l}}$} & \multicolumn{1}{c}{$\Delta_{\mathrm{u}}$} & & 
\multicolumn{1}{c}{$\log N$} & \multicolumn{1}{c}{$\Delta_{\mathrm{l}}$} & \multicolumn{1}{c}{$\Delta_{\mathrm{u}}$} &  & 
\multicolumn{1}{c}{$\log N$} & \multicolumn{1}{c}{$\Delta_{\mathrm{l}}$} & \multicolumn{1}{c}{$\Delta_{\mathrm{u}}$}\\
\multicolumn{1}{c}{(1)} & \multicolumn{1}{c}{(2)} & \multicolumn{1}{c}{(3)} & \multicolumn{1}{c}{(4)}&  &  \multicolumn{1}{c}{(5)}& \multicolumn{1}{c}{(6)}& \multicolumn{1}{c}{(7)} &  & \multicolumn{1}{c}{(8)} & \multicolumn{1}{c}{(9)} & \multicolumn{1}{c}{(10)} &  & \multicolumn{1}{c}{(11)} & \multicolumn{1}{c}{(12)} & \multicolumn{1}{c}{(13)} &  & \multicolumn{1}{c}{(14)}& \multicolumn{1}{c}{(15)}& \multicolumn{1}{c}{(16)}\\  \hline
16.25 & 2.79 & 0.09 & 0.09 & & 1.80 & 0.30 & 0.44 & & 2.11 & 0.21 & 0.26 & & 2.58 & 0.18 & 0.12 & & 3.08 & 0.10 & 0.08 \\
16.75 & 2.82 & 0.09 & 0.09 & & 2.11 & 0.21 & 0.26 & & 2.28 & 0.17 & 0.20 & & 2.76 & 0.15 & 0.11 & & 3.19 & 0.09 & 0.07 \\
17.25 & 3.15 & 0.06 & 0.06 & & 2.65 & 0.11 & 0.12 & & 2.58 & 0.12 & 0.13 & & 3.19 & 0.09 & 0.07 & & 3.58 & 0.06 & 0.05 \\
17.75 & 3.21 & 0.06 & 0.06 & & 2.98 & 0.07 & 0.07 & & 2.50 & 0.13 & 0.14 & & 3.42 & 0.07 & 0.06 & & 3.74 & 0.05 & 0.04 \\
18.25 & 3.11 & 0.06 & 0.06 & & 3.25 & 0.05 & 0.05 & & 2.65 & 0.11 & 0.12 & & 3.66 & 0.05 & 0.05 & & 3.91 & 0.04 & 0.04 \\
18.75$^\mathrm{\natural}$ & 3.09 & 0.06 & 0.08 & & 3.44 & 0.04 & 0.04 & & --      & --      & --      & & 3.83 & 0.04 & 0.04 & & 4.03 & 0.03 & 0.03 \\
\hline
\end{tabular}
\end{center}
} 
$^ \mathrm{\star}$ For the equivalence of the classes defined in P13 with traditional Hubble types, see \S\ref{sec:data}. \\
$^ \mathrm{\ast}$ $K$-band apparent magnitudes in the Vega system. \\
$^ \mathrm{\dag}$ The GNCs are provided in units of mag$^{-1}$ deg$^{-2}$. Their logarithmic values are listed.\\
$^ \mathrm{\ddag}$ The $\Delta_\mathrm{l}$ and $\Delta_\mathrm{u}$ errors are defined as $\Delta_\mathrm{l}
 \equiv \log(N) - \log(N-\sigma_\mathrm{l})$ and $\Delta_\mathrm{u} \equiv \log(N+\sigma_\mathrm{u}) - \log(N)$,
 where $\sigma_\mathrm{d}$ and $\sigma_\mathrm{u}$ correspond to the lower and upper errors due to statistical
 counting and classification uncertainties in each case. \\
$^ \mathrm{\natural}$ The data in this last magnitude bin are not complete in detection. 
\end{minipage}
\end{table*}

\section{GNC models}
\label{sec:model}

We have compared the results obtained  on GNCs in the $K$ band in Sect.\,\ref{sec:data} with the predictions of different evolutionary models, in order to shed some light on the evolutionary mechanisms responsible for the 
slope change at $K\sim 17.5$\,mag. The models used for comparison were the following:

\begin{enumerate}

\item a model assuming PLE for all galaxy types, including the E--S0's \citep{2010hsa5.conf..289E};

\item the \citet{2007MNRAS.376....2K} model (KW07 hereafter), based on the $\Lambda$CDM scenario of the Millenium simulation; 

\item the EM06 model, assuming a recent buildup of E--S0's at $z=1.5$ and moderate dust extinction at the initial 
evolutionary stages of these galaxies; and

\item the EM10 model, proposing a hierarchical scenario for the formation of massive E--S0's based on observational 
major merger rates up to $z\sim 1.5$.
\end{enumerate}

The two first ones are already known to not reproduce the slope change of the $K$-band GNCs
 \citep{2006MNRAS.366..858K,2007MNRAS.376....2K}, but they are used here as a reference. We have 
compared the expectations of these four basic models with our observational GNCs in \S\ref{sec:results}. 
Below, we provide a brief description of the basic assumptions and main characteristics of each model.

\subsection{PLE model for E--S0's since $z=3$}
\label{sec:PLE}

The PLE model assumes that the stellar populations of all galaxy types have just evolved according to an SFH 
characteristic for each type, but no number evolution is considered \citep{2010hsa5.conf..289E}. We have used the 
\texttt{NCMOD} code by \citet{1998PASP..110..291G} to predict the GNCs in different bands. The code starts from the 
local LFs of different galaxy types and evolves them backwards in time by considering three sources of evolution: the 
number evolution according to Gardner's implementation of merging (it is switched off in this case of PLE), the 
typical luminosity evolution of each galaxy type due to the SFH assumed for it (including k- and e-corrections), 
and the evolution of the volume element derived from the assumed cosmology. The `counter-evolved' LF of a given 
morphological type at each redshift is used to estimate the contribution of these galaxies to the total GNCs at a 
given magnitude and photometric band.

The galaxy types considered at $z\sim 0$ in the PLE model have been the same as those used in EM10 (E--S0, Sa--Sab, 
Sb--Sbc, Sc--Scd, and Sd--Irr). The Schechter functions of the LFs at $z=0$ for each type are described in table\,1 of EM10. The SFH of each galaxy type is fixed through the star formation rate (SFR) assigned to each type and the redshift at which the star formation activity in each galaxy class starts ($z_f$). In general, exponential functions have been used to represent the SFRs of spirals, instantaneous bursts for E--S0's, and constant SFRs for Sd-Irr and Im types. Different exponentially decaying timescales have been considered for spirals, according to observations \citep[$\tau = 4$, 7, and 9\,Gyr for Sa--Sab, Sb--Sbc, and Sc--Scd respectively, see][]{2005MNRAS.362...41G}. The characteristic metallicity of each type has also been fixed according to \citet{2005MNRAS.362...41G}. The complete set of parameters for describing the SFH and metallicity of each type can be found in table\,2 of EM10. The dust extinction for each type was set to typical observational values, as in the EM10 model (see \S\ref{sec:EM10} for more details). We have assumed $z_f=3$ for all types.

\subsection{Hierarchical assembly of E--S0's based on the Millenium Simulation (KW07 model)}
\label{sec:KW07}

In order to test a hierarchical scenario of galaxy formation, \citet[KW07 hereafter]{2007MNRAS.376....2K} compared the 
GNCs predicted in the $\Lambda$CDM cosmological framework of the Millenium simulation\footnote{Millenium Simulation Project 
webpage available at: http://www.mpa-\-garching.mpg.de/\-galform/\-virgo/\-millennium/.} with data in various bands. They 
derived the GNCs by constructing deep light-cone surveys from the attaching of semianalytic models of galaxy formation to the 
merging trees of the dark matter derived in the simulation \citep{2005Natur.435..629S}. The models accounted for the baryonic processes in the haloes and sub-haloes, including the `radio mode' feedback from the central galaxies in groups and clusters, and were very successful in reproducing several global properties of the galaxy population \citep{2006MNRAS.365...11C,2007MNRAS.375....2D}. 

KW07 already showed that this model reproduced the data of GNCs in optical bands, but predicted an overabundance of
massive galaxies at $z> 1$ that were too red, despite the hierarchical nature of the scenario \citep[see also][]{2009A&A...505.1041G}. We have compared our results with the total GNCs in several bands derived by these authors.

\subsection{Ad  hoc buildup of E--S0's at $z=1.5$ (EM06 model)}
\label{sec:EM06}

The EM06 model considered that the whole population of early-type galaxies appears on the cosmic scene at the same epoch ($z\sim 1.5$), but the model does not assume any mechanism for it. This framework is an obvious over-simplification of reality because the history of the specific SFR of present-day E--S0's points to a progressive build-up of their stellar populations, with star-formation peaks at some epochs depending on the mass \citep{2010MNRAS.404.1775T,2014ApJ...792...95C}. However, this model mimics a late structural assembly of E--S0's by approximating the redshift at which this build-up takes place by the redshift in which the bulk of their stars are born ($z_f=1.5$ in this model). 

The model is based on the original implementation of the \texttt{NCMOD} code (see \S\ref{sec:PLE} for a brief description). 
It assumes four galaxy types (E--S0, S0/a--Sb, Sbc--Sd, and Im), taken from the local LFs derived using SDSS data from
\citet{2003AJ....125.1682N}. Note that these are coarser than those used in the PLE and EM10 models (their Schechter 
parameterizations are listed in table\,11 of EM06). The redshift formation of all spiral types has been set to $z_f=3$, 
whereas $z_f=1.5$ for E--S0's. Similar SFHs and metallicities to those commented on in the PLE model were adopted in this
 model, including instantaneous starbursts for the E--S0 type (consult table\,12 in EM06).

The merging procedure implemented in the original \texttt{NCMOD} code was assumed but only affected the faint
 end of the GNCs (and that very negligibly), so we can neglect their effects here (see comments in EM06). The effects of dust 
extinction were already considered in the original \texttt{NCMOD} code. An average value of the $B$-band optical depth
 of $\tau_\mathrm{dust,star}=0.6$ was assumed for all morphological types (also for E--S0's). This value implies the
 consideration of strong dust extinction in the early phases of E--S0's, which was essential to avoid the appearance 
of a bump at faint magnitudes in the GNCs of blue optical bands caused by the blueish colours of these galaxies during the $\sim$$1$\,Gyr period after their build-up at $z=1.5$ (see \S\ref{sec:introduction}). As the EM06 model is not 
limited in magnitude range, it provides estimates on GNCs for all apparent magnitudes.

\subsection{Progressive hierarchical assembly of E--S0's (EM10 model)}
\label{sec:EM10}

In order to test the feasibility of the hierarchical origin of massive E--S0's ($M_\ast > 10^{11}\Msun$), EM10 studied 
how the population of massive early-type galaxies would evolve backwards in time under the assumption that they have 
derived from the major mergers that are observationally reported at each redshift up to $z\sim 1.5$ \citep[see
 also][]{2010arXiv1003.0686E}. Therefore, the EM10 model simulates a more realistic scenario for the build-up of massive 
E--S0's than the EM06 model, and distinguishes between the redshift of structural assembly of an E--S0 ($z_\mathrm{assembly}$) 
and the redshift at which the bulk of stars in the final galaxy formed in its progenitors ($z_f$).  In this model, the 
build-up of the local massive E--S0 population is progressive and can span a wide range of redshift values $z_\mathrm{assembly}$. 

The basic assumption of the model is that each major merger registered by observations has given rise to an E--S0 
galaxy, which is a reasonable hypothesis accounting for the typical gas contents of galaxies at $z<1.5$ (see 
references in EM10). The model accounts for the relative contribution of dry, mixed, and wet mergers, the 
existence of transient stages of major mergers with strong dust extinction and irregular morphologies (in which 
both galaxies have already merged into one body), and the typical timescales of these phases. 

The model predictions are limited to $M_B<-20$\,mag, which is the limiting magnitude of the observational sample
 used to derive galaxy merger rates. This limiting magnitude implies that the model can only trace back in time the evolution of E--S0's 
brighter than this magnitude at all redshifts because this is the only galaxy type assumed to be affected by the 
`inverse merging' of the model (i.e.\ it is the only one assumed to derive from major merging). This limits the 
model predictions to the E--S0 population that ends up with $M_\ast > 10^{11}\Msun$ at $z=0$ (see EM10 for more details).
 Therefore, the E--S0's sampled by the model at each redshift present lower masses as the model simulates the 
evolution backwards in time, not just because of the inverse SFH assumed for them (which is passive), but also because 
they include the gas-poor progenitors of the decomposed E--S0's, which obviously have lower masses than the remnant 
E--S0. The mass ratio of each encounter is randomly distributed between 1:1 and 3:1, meaning that the masses of the
 progenitors range from $\sim$$30$\% to $\sim$$60$\% of the stellar mass of the E--S0 resulting from the merger. 
The model distributes the estimated total number of major mergers occurring at each redshift according to the 
trends with the mass derived from observations and cosmological simulations \citep{2007MNRAS.375....2D,2009MNRAS.397.1776F}. The remaining Hubble types under consideration experience only luminosity evolution, except for the late-type discs, which also increase its population at the expense of the gas-rich progenitors of the E--S0's that are being decomposed at each redshift. 

The model was computed with an improved version of the \texttt{NCMOD} code. The local galaxy types considered were thinner than those assumed in the EM06 model (E--S0, Sa--Sab, Sb--Sbc, Sc--Scd, and Sd--Irr). The parameterizations of the LFs, SFHs, and metallicities of each type are the same as those described in the PLE model (\S\ref{sec:PLE}). All types have the redshift of start of their SFHs at $z_f=3$ (including the E--S0's).

In this model,which is strictly based on observations and on robust computational results, a more realistic treatment of the effects of major mergers on the LF of E--S0's was implemented. The model considered the different phases and time-scales in a major merger. Because the structural distortion and dust extinction of a merger remnant depend strongly on the characteristics of the progenitor galaxies, the model considered different properties of the E--S0 remnants during their transient relaxation phase, depending on whether they derive from a gas-rich or gas-poor encounters, according to the observed fraction of each merger type taking place at each redshift \citep{2008ApJ...681..232L}. The model assumed that the galaxies undergoing transient intermediate-to-advanced 
stages of major mergers had globally distorted morphologies and should be very red because of the strong dust 
reddening associated with the merger-induced starbursts in gas-rich events, or owing to the lack of star formation 
in gas-poor mergers. This is why the galaxies predicted to be undertaking an advanced stage of a gas-rich major 
merger were called dust-reddened star-forming galaxies (DSFs) in EM10. However, because we have dust-reddened starburst galaxies in our sample that are not necessarily related to major mergers, we will refer  more specifically in the model to these 
galaxies  as advanced stages of major mergers.  

The dust extinction procedure in the original {\tt NCMOD} code was improved by allowing the various galaxy types to acquire a different $\tau_\mathrm{dust,star}$ according to observational values (see table\,2 in EM10). The E--S0's were assumed to have little dust extinction, 
in better accordance with observations ($\tau_\mathrm{dust,star}=0.1$ in the $B$ band). For the intermediate-to-advanced stages of gas-rich major mergers, the model assumed a high dust extinction ($\tau_\mathrm{dust,star}=0.6$).

The limiting magnitude constrained by the observational merger rates basically traces the knee of the LF in the $K$ band (see fig.\,6 in EM10) and limits the model predictions of GNCs to $K\lesssim 18$\,mag. The EM10 model cannot therefore provide any estimates on the slope change in the $K$-band GNCs beyond $K\sim 18$\,mag by construction.  This limiting magnitude is equivalent to the following limits in other bands: $U\sim 25$, $B\sim 25$, $V\sim 25$, $R\sim 24.5$, $I\sim 24$, $z\sim 23.5$, $J\sim 20.5$, and $H\sim 19.5$\,mag. 
Nevertheless, we will show that it complements the simple scenario proposed in the EM06 model and provides very interesting clues to the nature of the NIR slope change when compared with the observational diagnostics derived in the present study. 

According to this model, the bulk of the local massive E--S0's underwent advanced stages of major merger events that assembled them at $0.6<z<1.2$, which means that these mergers started at $0.8<z<1.5$ approximately. Note that the EM06 and EM10 models share two key ingredients: 1) the massive E--S0's were definitively assembled in the universe at late epochs ($z\lesssim 1.5$), and 2) they had 
experienced stages of extreme dust extinction in the early stages of their assembly. However, the EM06 model does not assume any mechanisms to explain this build-up, whereas the EM10 directly tests whether major mergers since $z\sim 1.5$ could have built them up or not. We remark that the model can follow the evolution of Hubble types later than E--S0's at all masses, but only of E--S0's which have $M_\ast > 10^{11}\Msun$ at $z=0$.

\begin{table*}
 \begin{minipage}{\textwidth}
{\footnotesize
  \caption{$K$-band GNCs of red regular spheroid-dominated galaxies (E--S0's) in narrow redshift bins\label{Tab:GNCsES0}}
\begin{center}
  \begin{tabular}{lrrrcrrrcrrrcrrrcrrr}
  \hline
 & \multicolumn{3}{c}{$0.3<z<1.5$} & & \multicolumn{3}{c}{$0.4<z<0.6$} & & \multicolumn{3}{c}{$0.6<z<0.8$} & & \multicolumn{3}{c}{$0.8<z<1.0$} & & \multicolumn{3}{c}{$ 1.0<z<1.2$}\\\cline{2-4}\cline{6-8}\cline{10-12}\cline{14-16}\cline{18-20}\\\vspace{-0.5cm}\\
\multicolumn{1}{c}{$K^\mathrm{\ast}$} & \multicolumn{1}{c}{$\log N^\mathrm{\dag}$} & \multicolumn{1}{c}{$\Delta_{\mathrm{l}}^\mathrm{\ddag}$} & \multicolumn{1}{c}{$\Delta_{\mathrm{u}}^\mathrm{\ddag}$} & & 
\multicolumn{1}{c}{$\log N$} & \multicolumn{1}{c}{$\Delta_{\mathrm{l}}$} & \multicolumn{1}{c}{$\Delta_{\mathrm{u}}$} &  & 
\multicolumn{1}{c}{$\log N$} & \multicolumn{1}{c}{$\Delta_{\mathrm{l}}$} & \multicolumn{1}{c}{$\Delta_{\mathrm{u}}$} & & 
\multicolumn{1}{c}{$\log N$} & \multicolumn{1}{c}{$\Delta_{\mathrm{l}}$} & \multicolumn{1}{c}{$\Delta_{\mathrm{u}}$} &  & 
\multicolumn{1}{c}{$\log N$} & \multicolumn{1}{c}{$\Delta_{\mathrm{l}}$} & \multicolumn{1}{c}{$\Delta_{\mathrm{u}}$}\\
\multicolumn{1}{c}{(1)} & \multicolumn{1}{c}{(2)} & \multicolumn{1}{c}{(3)} & \multicolumn{1}{c}{(4)}&  &  \multicolumn{1}{c}{(5)}& \multicolumn{1}{c}{(6)}& \multicolumn{1}{c}{(7)} &  & \multicolumn{1}{c}{(8)} & \multicolumn{1}{c}{(9)} & \multicolumn{1}{c}{(10)} &  & \multicolumn{1}{c}{(11)} & \multicolumn{1}{c}{(12)} & \multicolumn{1}{c}{(13)} &  & \multicolumn{1}{c}{(14)}& \multicolumn{1}{c}{(15)}& \multicolumn{1}{c}{(16)}\\  \hline
16.25 &   2.79 &    0.22 &   0.22 &  & 2.41 & 0.32 &  0.32 & & 1.93 &  0.56 &  0.52 & & 2.11 &  0.51 &  0.43 & &   --   &   --   &  --   \\
16.75 &   2.82 &    0.21 &   0.21 &  & 2.23 & 0.41 &  0.38 & & 2.53 &  0.32 &  0.28 & & 1.81 &  0.83 &  0.58 & &   --   &   --   &  --   \\
17.25 &   3.15 &    0.15 &   0.17 &  & 2.63 & 0.25 &  0.26 & & 2.71 &  0.24 &  0.24 & & 2.11 &  0.41 &  0.44 & &  2.33 & 0.48 & 0.35 \\
17.75 &   3.21 &    0.14 &   0.16 &  & 2.37 & 0.31 &  0.34 & & 2.90 &  0.21 &  0.21 & & 2.69 &  0.24 &  0.24 & &  1.63 & 0.74 & 0.70 \\
18.25 &   3.11 &    0.16 &   0.17 &  & 2.53 & 0.28 &  0.28 & & 2.44 &  0.33 &  0.31 & & 2.50 &  0.26 &  0.30 & &  2.47 & 0.36 & 0.29 \\
18.75$^ \mathrm{\natural}$ &   3.09 &    0.16 &   0.17 &  & 1.63 & 0.70 &  0.72 & & 2.73 &  0.32 &  0.23 & & 2.58 &  0.25 &  0.27 & &  2.17 & 0.53 & 0.41 \\
\hline
\end{tabular}
\end{center}
} 
$^ \mathrm{\ast}$ $K$-band apparent magnitudes in the Vega system. \\
$^ \mathrm{\dag}$ The GNCs are provided in units of mag$^{-1}$ deg$^{-2}$. Their logarithmic values are listed.\\
$^ \mathrm{\ddag}$ The $\Delta_\mathrm{l}$ and $\Delta_\mathrm{u}$ errors are defined as
 $\Delta_\mathrm{l} \equiv \log(N) - \log(N-\sigma_\mathrm{l})$ and
 $\Delta_\mathrm{u} \equiv \log(N+\sigma_\mathrm{u}) - \log(N)$, where $\sigma_\mathrm{d}$
 and $\sigma_\mathrm{u}$ correspond to the lower and upper errors due to statistical counting,
 classification, and cosmic variance uncertainties in each case.\\
$^ \mathrm{\natural}$  The data in this last magnitude bin are not complete in detection. 
\end{minipage}
\end{table*}

\begin{table*}
 \begin{minipage}{\textwidth}
{\footnotesize
  \caption{$K$-band GNCs of red regular spheroid-dominated galaxies (E--S0's) in wide overlapping redshift bins\label{Tab:GNCsES02}}
\begin{center}
  \begin{tabular}{lrrrcrrrcrrrcrrr}
  \hline
 & \multicolumn{3}{c}{$0.3<z<1.5$} & & \multicolumn{3}{c}{$0.4<z<0.8$} & & \multicolumn{3}{c}{$0.6<z<1.0$} & & \multicolumn{3}{c}{$0.8<z<1.2$} \\\cline{2-4}\cline{6-8}\cline{10-12}\cline{14-16}\\\vspace{-0.5cm}\\
\multicolumn{1}{c}{$K^\mathrm{\ast}$} & \multicolumn{1}{c}{$\log N^\mathrm{\dag}$} & \multicolumn{1}{c}{$\Delta_{\mathrm{l}}^\mathrm{\ddag}$} & \multicolumn{1}{c}{$\Delta_{\mathrm{u}}^\mathrm{\ddag}$} & & 
\multicolumn{1}{c}{$\log N$} & \multicolumn{1}{c}{$\Delta_{\mathrm{l}}$} & \multicolumn{1}{c}{$\Delta_{\mathrm{u}}$} &  & 
\multicolumn{1}{c}{$\log N$} & \multicolumn{1}{c}{$\Delta_{\mathrm{l}}$} & \multicolumn{1}{c}{$\Delta_{\mathrm{u}}$} & & 
\multicolumn{1}{c}{$\log N$} & \multicolumn{1}{c}{$\Delta_{\mathrm{l}}$} & \multicolumn{1}{c}{$\Delta_{\mathrm{u}}$} \\
\multicolumn{1}{c}{(1)} & \multicolumn{1}{c}{(2)} & \multicolumn{1}{c}{(3)} & \multicolumn{1}{c}{(4)}&  &  \multicolumn{1}{c}{(5)}& \multicolumn{1}{c}{(6)}& \multicolumn{1}{c}{(7)} &  & \multicolumn{1}{c}{(8)} & \multicolumn{1}{c}{(9)} & \multicolumn{1}{c}{(10)} &  & \multicolumn{1}{c}{(11)} & \multicolumn{1}{c}{(12)} & \multicolumn{1}{c}{(13)} \\  \hline
16.25  & 2.80  & 0.27 & 0.20&   & 2.53  & 0.35 & 0.26&   & 2.33  & 0.46  & 0.32&  & 2.17  & 0.57  & 0.37\\
16.75  & 2.82  & 0.26 & 0.20&   & 2.71  & 0.29 & 0.22&   & 2.61  & 0.32  & 0.24&  & 1.81  & 1.67  & 0.54\\
17.25  & 3.15  & 0.17 & 0.16&   & 2.97  & 0.21 & 0.18&   & 2.80  & 0.23  & 0.21&  & 2.50  & 0.29  & 0.28\\
17.75  & 3.21  & 0.17 & 0.15&   & 3.01  & 0.19 & 0.18&   & 3.11  & 0.19  & 0.16&  & 2.73  & 0.28  & 0.22\\
18.25  & 3.11  & 0.19 & 0.16&   & 2.79  & 0.26 & 0.21&   & 2.78  & 0.25  & 0.21&  & 2.79  & 0.26  & 0.20\\
18.75$^ \mathrm{\natural}$  & 3.10  & 0.20 & 0.16&   & 2.77  & 0.28 & 0.20&   & 2.96  & 0.22  & 0.18&  & 2.73  & 0.25  & 0.22\\
\hline
\end{tabular}
\end{center}
} 
$^ \mathrm{\ast}$ $K$-band apparent magnitudes in the Vega system. \\
$^ \mathrm{\dag}$ The GNCs are provided in units of mag$^{-1}$ deg$^{-2}$. Their logarithmic values are listed.\\
$^ \mathrm{\ddag}$ The $\Delta_\mathrm{l}$ and $\Delta_\mathrm{u}$ errors are defined as $\Delta_\mathrm{l}
 \equiv \log(N) - \log(N-\sigma_\mathrm{l})$ and $\Delta_\mathrm{u} \equiv \log(N+\sigma_\mathrm{u}) - \log(N)$,
 where $\sigma_\mathrm{d}$ and $\sigma_\mathrm{u}$ correspond to the lower and upper errors due to statistical counting,
 classification, and cosmic variance uncertainties in each case.\\
$^ \mathrm{\natural}$  The data in this last magnitude bin are not complete in detection.
\end{minipage}
\end{table*}

\section{Results and discussion}
\label{sec:results}

In \S\ref{sec:GNCsobserved}, we first comment on the results concerning the nature of the $K$-band slope 
change that can be directly derived from the observed GNCs by morphological types and of E--S0's 
by redshift bins obtained in this study. We compare these and other GNC data from the literature with the 
predictions of different galaxy evolutionary models in \S\ref{sec:gncspredicted}.

\subsection{Observational results}
\label{sec:GNCsobserved}

\subsubsection{GNCs by galaxy types in the $K$ band}
\label{sec:counts}

\begin{figure}
\includegraphics[width=0.5\textwidth,angle=0, bb= 12 10 340 340, clip]{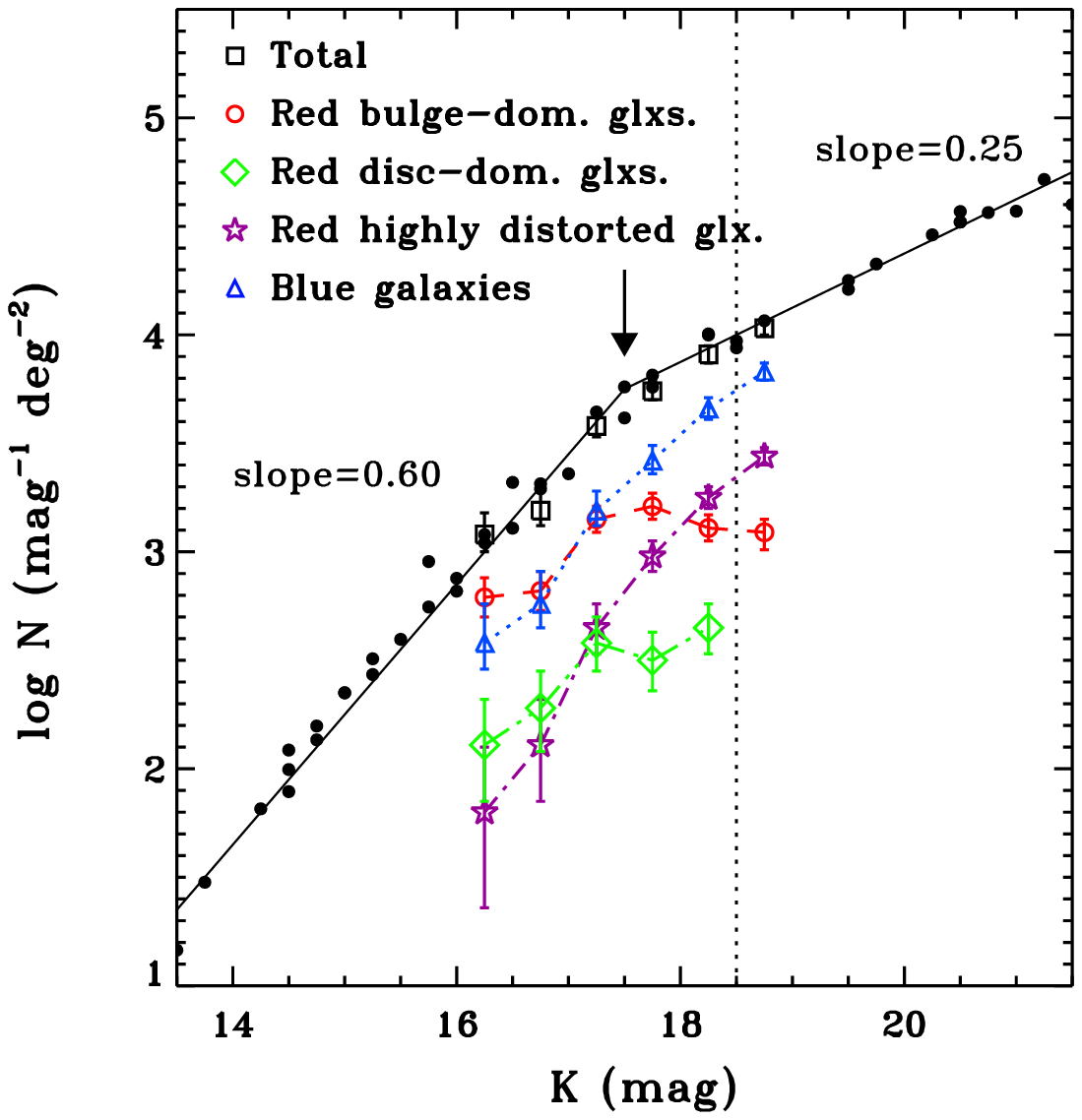}
\caption{Observational GNCs in the $K$ band obtained in this study for all galaxies (total), E--S0's, 
blue galaxies, red highly distorted galaxies, and red regular discs (see the legend in the figure). 
\emph{Vertical dotted line}: limiting magnitude for 100\% completeness of our data. \emph{Black filled circles}: complete GNCs dataset compiled by CH03 in the $K$ band from the literature, plotted for comparison 
\citep[][CH03]{1993ApJ...415L...9G,1995ApJ...438L..13D,1995ApJS...96..117M,1997ApJ...480..503H,1998ApJ...494..111M,1998ApJ...492..452S,2001ApJ...559..592T}. \emph{Solid lines}: fits performed to the bright and faint ends of the GNCs by CH03. The slopes obtained by these authors are indicated in the figure. \emph{Black arrow}: location of the slope change in total GNCs.  [\emph{A colour version is available at the online version.}]}\label{Fig:cuentas}
\end{figure}

In Fig.\,\ref{Fig:cuentas}, we represent the GNCs in the $K$ band obtained in the present study for total, E--S0's, 
red highly distorted galaxies, red regular discs, and blue galaxies (see Table\,\ref{Tab:GNCs}). We have 
overplotted the GNC compilation of complete data performed in the band by CH03, in order to check the reliability 
of our results by comparing with independent studies. Our total GNCs computed for galaxies at $0.3<z<1.5$ agree 
pretty well with the data obtained by other authors for magnitudes brighter than our completeness limit 
($K\sim 18.5$\,mag) down to $K\sim 16$\,mag, where our sample starts to suffer from poor statistics owing to 
its area. Therefore, the figure indicates that the total GNCs at $16<K<18.5$\,mag are made up by the galaxy 
populations at $0.3<z<1.5$.

The figure also shows that the galaxy types that experience a change in the slope of their GNCs around 
$K\sim 17.5$\,mag are E--S0's and red discs. However, the contribution of red discs to the total GNCs is
 negligible at these magnitudes, meaning that the slope change in total $K$-band GNCs is caused by a change 
in the trend of the GNCs of E--S0 galaxies. This decrease in the contribution of E--S0's to the total GNCs at $K\sim 17.5$\,mag contrasts with the rising contributions of blue galaxies and red distorted galaxies, which preserve their rising slope up to $K\sim 18.5$\,mag. 

Figure\,\ref{Fig:cuentas} demonstrates that the slope change at $K\sim 17.5$\,mag is due to a change of the
galaxy population that numerically dominates the total $K$-band GNCs: from the quiescent regular E--S0's at
 $K< 17.5$\,mag to blue discs at $K> 17.5$\,mag. This is the first direct observational evidence for the fact
 that E--S0's are the galaxy population responsible for the slope change (as proposed by so many studies, 
see \S\ref{sec:introduction}), because the slope of the total GNCs decreases at this magnitude owing to a
 flattening of the contribution of E--S0's at $K>17.5$\,mag (which slightly decreases, instead of continuing to rise, as observed at brighter magnitudes). 

\subsubsection{GNCs of E--S0's by redshift bins in the $K$ band}
\label{sec:countsES0}

\begin{figure}
\includegraphics[width=0.45\textwidth,angle=0, bb= 12 10 340 340, clip]{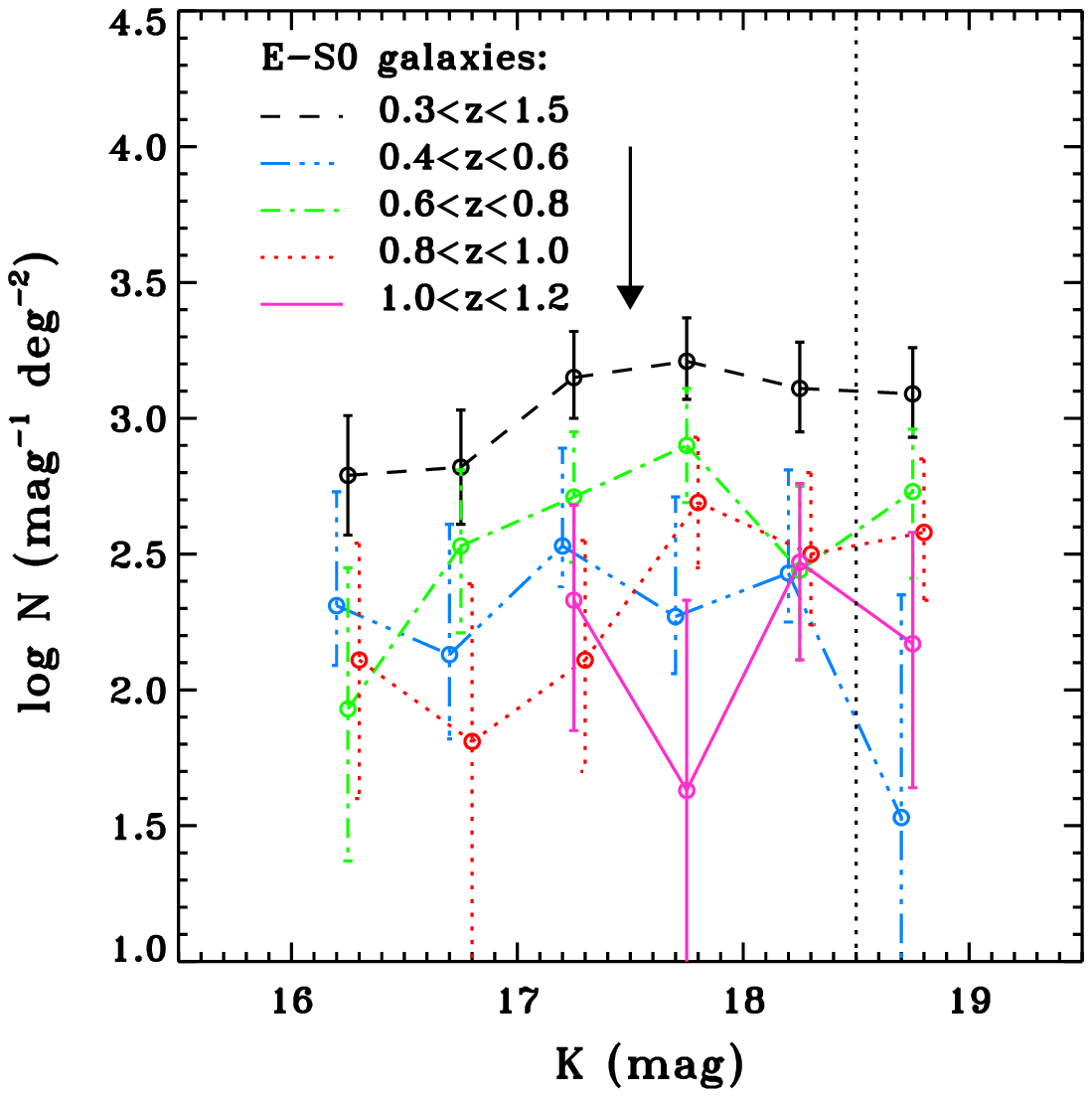}
\includegraphics[width=0.45\textwidth,angle=0, bb= 12 10 340 340, clip]{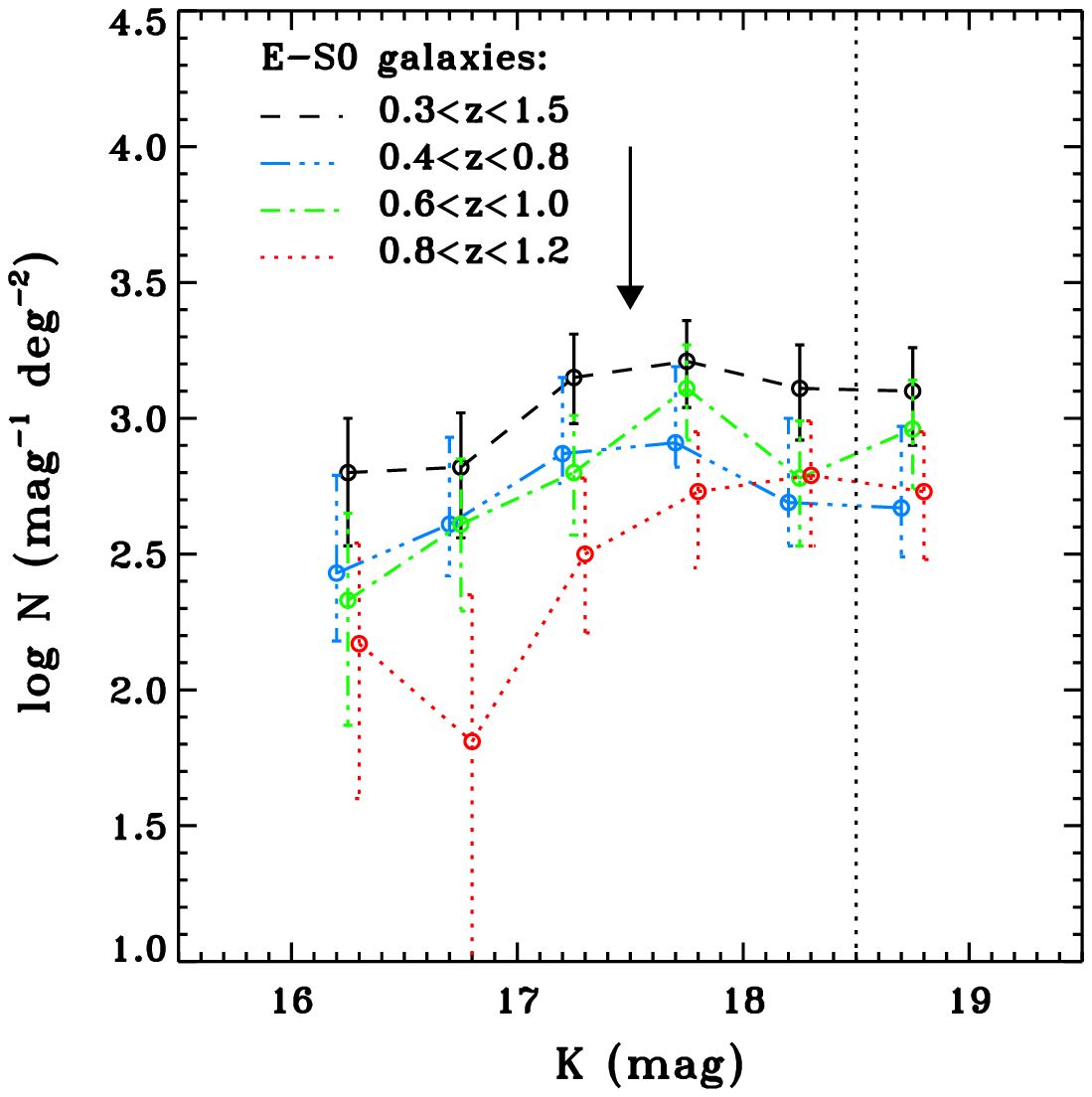}
\caption{Observational $K$-band GNCs of E--S0 galaxies obtained in the present study for different
 redshift bins up to $z<1.2$ (see the legend in the figure). The total GNCs of E--S0's in the band
 for $0.3<z<1.5$ from Fig.\,\ref{Fig:cuentas} have been overplotted for reference. \emph{Top panel}:
 for narrow redshift bins. \emph{Bottom panel}: for wide, overlapping redshift bins. \emph{Vertical 
dotted line}: limiting magnitude for 100\% completeness of our data. \emph{Black arrow}: location of
 the slope change in the total GNCs.  [\emph{A colour version is available in the online version.}]}
\label{Fig:cuentasEs}
\end{figure}

In order to find out the range of redshifts of the E--S0's responsible for the slope change, we have represented the $K$-band GNCs of E--S0's by redshift bins in Fig.\,\ref{Fig:cuentasEs}. The top panel shows the distributions for narrow redshift bins (data in Table\,\ref{Tab:GNCsES0}), while the bottom panel represents the results obtained for the three wide overlapping redshift bins 
(see Table\,\ref{Tab:GNCsES02}). We have overplotted the total contribution of E--S0's at $0.3<z<1.5$ 
from Fig.\,\ref{Fig:cuentas} as a reference in both panels. The large error bars in the top panel 
prevent us from deriving any  conclusions. However, the GNCs of E--S0's in the wider redshift bins of the 
bottom panel have more reasonable errors that enable us to derive some conclusions.

The bottom panel of Fig.\,\ref{Fig:cuentasEs} shows that the slope of the total GNCs of E--S0's at
 bright magnitudes decreases at $K\sim 17.5$\,mag because the contribution of these galaxies flattens or 
decreases at $K>17$ for all redshift bins. The E--S0's that should dominate the GNCs at fainter magnitudes
 according to the extrapolations of their trends at brighter magnitudes are those located at $0.6<z<1.0$ and
 $0.8<z<1.2$. However, their contributions decrease instead of contributing to raise the GNCs at fainter
 magnitudes. This is consistent with the results by \citet{2009A&A...494...63B}, who claimed that the
 galaxies responsible for the slope change were at $z<1$ (although these authors did not identify the
 type of galaxies). We will show that these GNCs of E--S0's by redshift bins provide evidence of a
 noticeable number evolution of massive E--S0's at $z>0.6$ when compared with the models (see \S\ref{sec:gncsES0zbins}).

In conclusion, the $K$-band morphological GNCs for different galaxy types and the GNCs for E--S0's 
in redshift bins evidence that the slope change in the NIR GNCs is caused by a change in the contribution of E--S0's at $0.6<z<1.2$ with respect to the expectations derived from the extrapolations of their trends at brighter magnitudes. We note that many studies had previously proposed this fact  (see \S\ref{sec:introduction}), but this is the first time that it is shown directly with observational data.

\subsection{Clues derived from the comparison with evolutionary models}
\label{sec:gncspredicted}

In this section, we compare the GNC data derived in the present and other studies with the predictions of the models commented on in \S\ref{sec:model}. We show that a number of conclusions can be derived in this way concerning how rapidly the massive end of the red sequence was established during the last $\sim 9$\,Gyr.

\subsubsection{\bf Total GNCs from the NUV to the NIR: evidence of a relatively late definitive build-up of $L\sim L^\ast$ E--S0's}
\label{sec:gncsmuultiwavelength}

\begin{figure*}
\begin{center}
\includegraphics*[width=0.95\textwidth,angle=0]{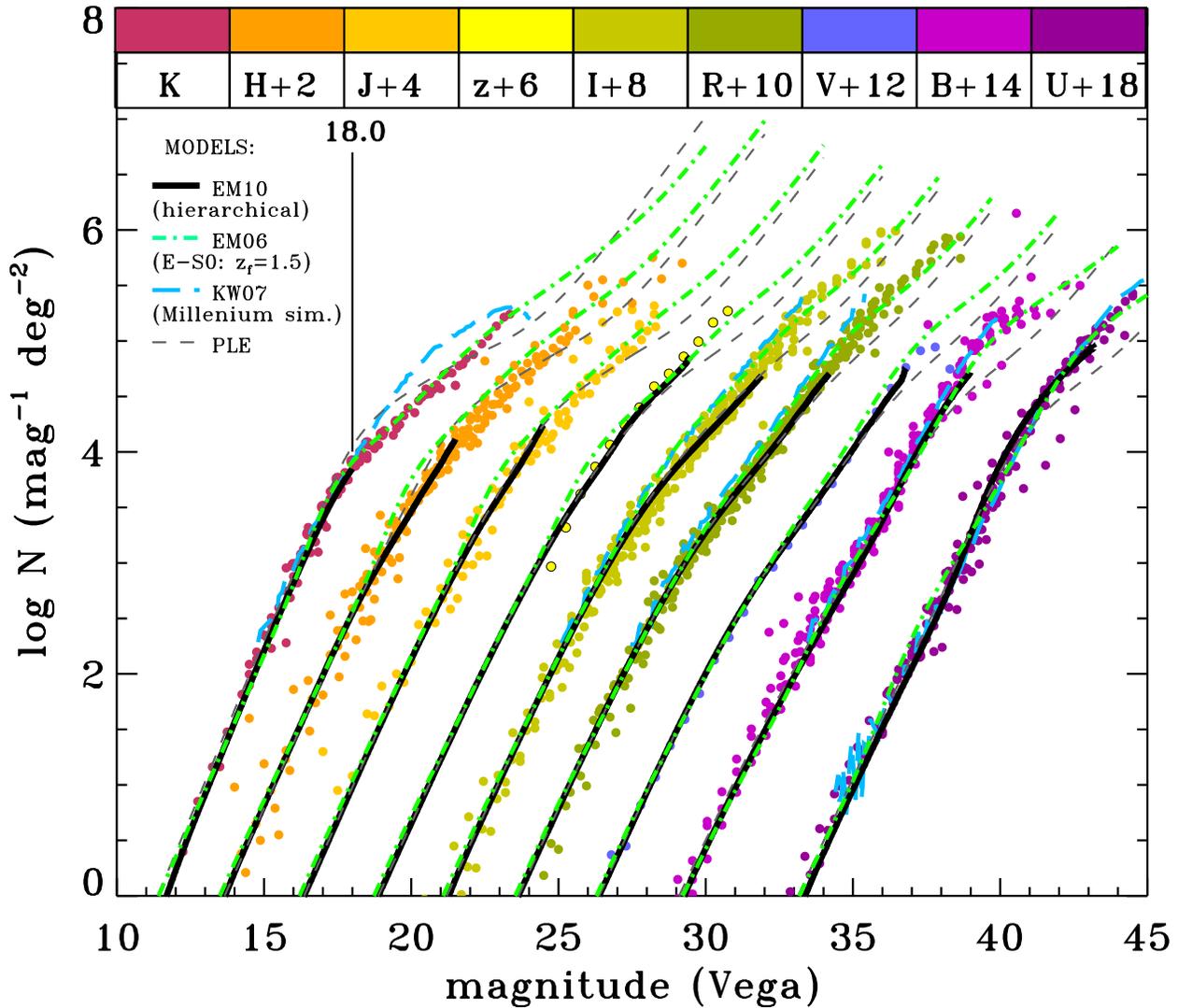}
\caption{ Predictions of the models on the total GNCs from the NUV to the NIR bands compared with real data 
(\emph{coloured circles}). The GNCs of each band has been displaced in the magnitude axis by a constant for
 clarity, as shown in the legend of data colours at the top. \emph{Black solid lines}: predictions 
of the EM10 model in all bands. \emph{Green dotted-dashed lines}: predictions of the EM06 model in all bands.
 \emph{Light-blue long-dashed lines}: predictions of the KW07 model in the $U$, $B$, $R$, $I$, and $K$ bands.
 \emph{Grey dashed lines}: predictions of the PLE model for all bands.\/ This figure uses the compilation 
of published GNC data by N.~Metcalfe in the $R$, $I$, and $H$ bands based on CCD imaging. The GNCs for the $K$
 band include our  total GNCs, the selection of complete data performed by CH03, and the deep data by 
\citet{2010A&A...511A..50R}. \newline\newline
The references for each band are the following: 
\textbf{\emph{U} band} \citep{1990ApJ...348..371S,
2001A&A...379..740A,2001AJ....122.1104Y,2004AJ....127..180C,2004A&A...417...51R,2006ApJ...639..644E,
2009A&A...505.1041G,2009ApJ...705.1462H,2009A&A...507..195R}, \textbf{\emph{B} band} \citep{2001A&A...379..740A,
2001A&A...368..787H,2001A&A...370..384K,2001AJ....122.1104Y,2003A&A...410...17M,2004AJ....127..180C,
2004PASJ...56.1011K,2006ApJ...639..644E,2009A&A...507..195R}, \textbf{\emph{V} band}\citep{1996MNRAS.282L...1G,2004AJ....127..180C}, \textbf{\emph{R} band} \citep{1984ApJS...56..143C,
1984MNRAS.210..979H,1985PhDT.......217S,1986AJ.....91..217I,1986ApJ...311..651K,1987ApJ...319...28Y,
1988AJ.....96....1T,1991MNRAS.249..481J,1991MNRAS.249..498M,1991AJ....102..445P,1993MNRAS.260..241C,
1993AJ....105.2017S,1994MNRAS.266..155D,1995MNRAS.273..257M,1995ApJ...449L.105S,1997A&A...317...43B,
1997MNRAS.288..404H,1998MNRAS.294..147M,1999A&A...341..641A,2001A&A...379..740A,2001A&A...368..787H,
2001A&A...370..384K,2001MNRAS.323..795M,2003A&A...410...17M,2004AJ....127..180C,2004PASJ...56.1011K}, 
\textbf{\emph{I} band} \citep{1984MNRAS.210..979H,1986ApJ...311..651K,1988AJ.....96....1T,
1991ApJ...369...79L,1994MNRAS.266..155D,1995ApJ...453..599C,1995ApJ...449L..23D,1995MNRAS.275L..19G,
1995ApJ...455...60L,1995ApJ...449L.105S,1996MNRAS.282L...1G,1998ApJ...506...33P,1999A&A...341..641A,
2001A&A...379..740A,2001MNRAS.323..795M,2001AJ....122.1104Y,2003A&A...410...17M,2004AJ....127..180C,
2007ApJS..172...99C,2004PASJ...56.1011K}, \textbf{\emph{z} band} \citep{2004AJ....127..180C}, 
\textbf{\emph{J} band} \citep{1998ApJ...505...50B,2000ApJ...540..593V,2001A&A...375....1S,2007AJ....133.2418I,
2009ApJ...696.1554C,2010A&A...511A..50R}, \textbf{\emph{H} band} \citep{1998ApJ...507L..17T,1998ApJ...503L..19Y,
1999AJ....117...17T,2001AJ....121..598M,2001MNRAS.323..795M,2002ApJ...570...54C,2006MNRAS.371.1601F,
2009ApJ...696.1554C,2010ApJS..186...94K,2010A&A...511A..50R}, \textbf{\emph{K} band} \citep[][ results derived in
 the present study, and references in the caption of Fig.\,\ref{Fig:cuentas}]{2010A&A...511A..50R}.  [\emph{A 
colour version is available in the online version.}]
} \label{fig:ncs}
\end{center}
\end{figure*}

Figure\,\ref{fig:ncs} compares GNC data in several broad bands from the NUV to the NIR together with the predictions 
of the four evolutionary scenarios indicated in \S\ref{sec:model}. The predictions of 
the PLE, EM06, and EM10 models are plotted for all bands. The KW07 model is plotted only for the $U$, 
$B$, $R$, $I$, and $K$ filters for clarity. The figure uses the compilation of published GNCs 
performed by N.\,Metcalfe\footnote{Galaxy number counts page by the Extragalactic Astronomy and Cosmology
 Research Group of the Durham University: http://star-www.dur.ac.uk/~nm/pubhtml/counts/counts.html} for
 the $R$, $I$, and $H$ bands and for several fields. The data and the equations used to convert between
 bands and the AB and Vega systems are detailed in his webpage. In the $K$ band, we have overplotted the 
total GNCs obtained in this study, as well as the dataset of complete GNCs data compiled by CH03. The
 selection of complete data reduces the typical dispersion between authors owing to different extraction
 procedures, differences in the filters, and/or cosmic variance between the fields, which usually smudge
 the slope change at $K\sim 17.5$\,mag. We have also represented the deep GNCs obtained by 
\citet{2010A&A...511A..50R} in UDF and GOODS-South fields in the $J$, $H$, and $K$ bands. References for 
all data in the figure are provided in the caption.

The PLE model cannot fit the GNC data at faint magnitudes in any band, as shown in Fig.\,\ref{fig:ncs}. 
The predictions of PLE are above the data in the $K$ band and below them in blue optical filters. In particular,
 the PLE model rolls over at $K\sim 20$\,magnitude, basically because of the evolution of the cosmological volume element, the change 
in the k-corrections in the $K$ band from being positive to negative, and the different shapes of the LFs representing E--S0's and spirals \citep{2003RMxAC..16..203B}. However, this slope change appears to be $\sim$$2.5$\,mag deeper than the observational slope change at $K\sim 17.5$\,mag. This mismatch indicates that the PLE model predicts too few blue galaxies and too many red ones at high redshifts, 
which clearly indicates the existence of a conversion of blue into red galaxies in the universe (see references of previous studies already stating this in \S\ref{sec:introduction}). 

The KW07 model fits the observations pretty well in the $U$, $B$, $R$, and $I$ bands, but clearly fails to
 reproduce total GNCs in the $K$ band faintwards of $K\sim 19$\,mag. This discrepancy was already noticed by
 these authors, who concluded that the evolutionary scenario of the Millenium simulation `overpredicts the 
abundance of moderately massive galaxies at high redshift, despite the fact that late merging plays a major
 role in the build-up of its more massive galaxies'. This statement means that massive galaxies are being assembled 
at earlier times in the model (in fact, at $z>2$) than in the real universe. This model may be missing some 
physical mechanisms that delay the definitive major assembly of these galaxies until later epochs (at $z<1.5$), 
according to the total GNCs from NUV to NIR bands. \citet{2009A&A...505.1041G} also compared deep $U$ 
and $K$ GNCs with the predictions of three hierarchical CDM models (including also the KW07 model, see their
 figs.\,3 and 4). They found large discrepancies at the faint end of the $U$-band GNCs between the three models that
clearly favour the KW07 model. They attributed these discrepancies to differences between the models in the recipes dealing with  the star 
formation activity and dust extinction of the faint galaxy population at $z<2$.

The EM06 and EM10 models fit well all the data from NUV to NIR in their ranges of validity. This goodness of fit suggests that
 they may be describing the build-up of the bulk of the E--S0's more realistically than the PLE and KW07
 models in global terms. In particular, the EM06 model, besides reproducing the slope change at 
$K\sim 17.5$\,mag, can also predict the faint end of the GNCs in the $K$ band up to $K\sim 24$\,mag drawn
 by the data by \citet{2010A&A...511A..50R}. These authors reported that they found no change in the slope 
of the $K$ GNCs at this very faint end (i.e.\ for $K>20$\,mag), so it is extremely significant that the
 EM06 model reproduces it, whereas the PLE model cannot. Both the PLE and EM06 models assume PLE for all 
Hubble types later than Sa according to standard SFHs and very high formation redshifts ($z_f=3$), so, in this sense, they 
are analogous. They differ basically  in that the PLE model assumes that E--S0's appear on 
the cosmic scene at $z_f=3$, whereas the EM06 model delays their appearance until $z_f=1.5$. Therefore, the shape of 
the total $K$-band GNCs directly rejects passive evolutionary scenarios for  E--S0's from very high redshifts
 not just because of the slope change at $K\sim 17.5$\,mag, but also because of their featureless slope at the 
very faint end. 

Although the EM10 has a much shorter magnitude range of validity (see \S\ref{sec:EM10}), it nearly 
follows the predictions of the EM06 model, in particular, the slope change until $K=18$\,mag. We note that the EM06 and EM10 models differ from the PLE and KW07 models in that they propose a late build-up of the bulk of the massive early type galaxies (at $z\lesssim1.5$). The EM06 model imposed it ad hoc, whereas the EM10 model derived it naturally from the major mergers registered up to
 $z\sim 1.5$ (assuming that one E--S0 has risen from each observed major merger). In this sense, the 
EM06 model can be considered a coarser approximation to the scenario proposed in the EM10 model, but with 
the clear advantage that it can make predictions at all magnitudes, whereas the EM10 model is limited to
 $M_B<-20$\,mag. Therefore, we can complement the predictions of the EM10 model by those of the EM06 model at very faint magnitudes, considering the latter to be a rough sketch of the formation scenario proposed in the former.

Note also that the EM06 and EM10 models are capable of reproducing $U$-band GNCs down to $U\sim 26$--27\,mag at least as well as the KW07 model, whereas the PLE model fails to predict them at the faint end. Therefore, models assuming PLE evolution of E--S0's since very high redshifts can be neglected not only by attending to the GNCs in the $K$ band, but also to the deep GNCs in the  optical-to-NUV bands.

\begin{figure}
\begin{center}
\includegraphics[width=0.5\textwidth,angle=0]{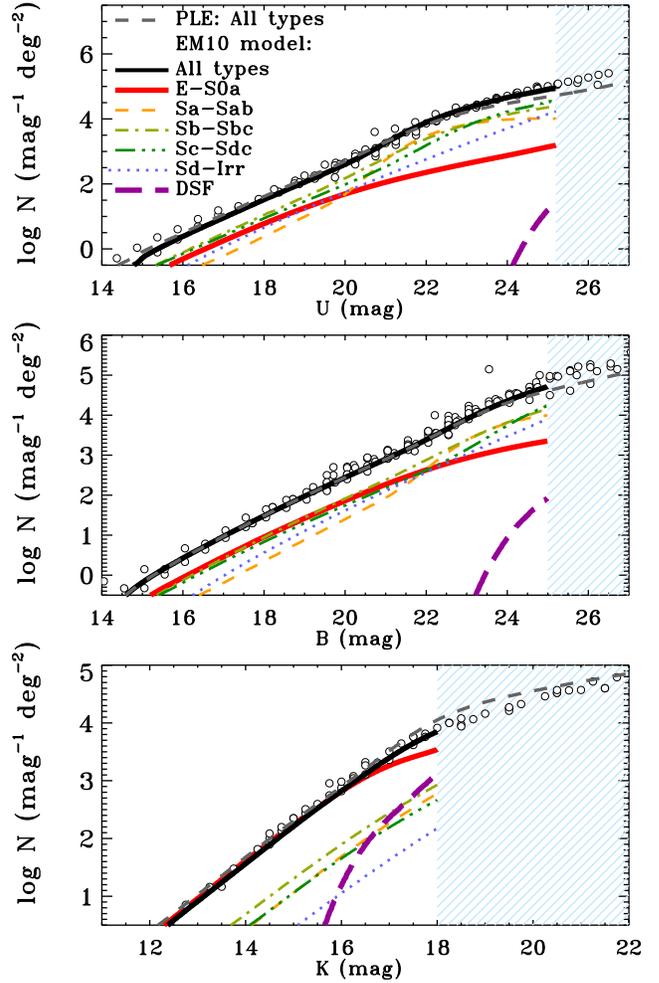}
\caption{EM10 model predictions of the contribution of the various morphological types to the total 
GNCs in the $U$, $B$, and $K$ bands compared to observational data (see references in the caption of 
Fig.\,\ref{fig:ncs}). The region where the predictions of the EM10 model are not valid in each band is
 shaded in light blue. \emph{Black thick solid lines}: total GNCs predicted by the EM10 model in each
 band. \emph{Grey dashed lines}: predictions of the PLE model on the total GNCs in each band, plotted as 
a reference. \emph{Rest of lines}: EM10 model predictions of the GNCs in each band due to the E--S0's
 (\emph{red solid}), Sa-Sab's (\emph{yellow dashed}), Sb-Sbc's (\emph{olive green dotted-dashed}), Sc-Scd's
 (\emph{dark green three dotted-dashed}), Sd-Irr's (\emph{blue dotted}), and red advanced stages of
 major mergers  (\emph{purple thick long-dashed}).  [\emph{A colour version is available at the online version.}] } \label{fig:ncstypesK2}
\end{center}
\end{figure}

Figure\,\ref{fig:ncstypesK2} shows the contribution of the various galaxy types to the total GNCs predicted by the EM10 model in the $U$, $B$, and $K$ bands, compared with the same observational data as in Fig.\,\ref{fig:ncs}. The model is capable of reproducing the slope change until $K\sim 18$\,mag (the limiting magnitude of the model) and attributes it to a flattening of the contribution 
of E--S0's to the total GNCs at $K\gtrsim 16$\,mag. The GNCs of the later types (from Sa's to Sd's) do not 
significantly change the rising trend they exhibit at bright magnitudes after the break. The figure 
shows that $K$-band GNCs are controlled by E--S0's up to the slope change. Accounting for the shape of their 
LF up to $z\sim 2$ \citep{2007A&A...476..137A,2007MNRAS.380..585C}, NIR GNCs are controlled by $L>L^{\ast}$ E--S0's up to $K\sim 17.5$\,mag according to this model. We will also show that the addition of the GNCs from spiral types in the model exceeds the GNCs of E--S0's after the break, so the global scenario proposed by the EM10 model is in good agreement with the observational results that we have derived in \S\ref{sec:GNCsobserved} on the nature of the slope change in the $K$-band GNCs (see \S\ref{sec:gncsmorphinbands}). 
 
To summarize, the observed distributions of total GNCs from the NUV to the NIR bands support a late definitive build-up 
for the majority of the E--S0 population (at $z<1.5$) against the scenarios proposing that they have evolved 
passively since very early epochs (at $z>2$). This late epoch of assembly of massive E--S0's (postulated by the EM06 model and predicted by the EM10 model under the assumption that each observed major merger up to $z\sim 1.5$ has produced an E--S0 
galaxy)  is in good agreement with traditional hierarchical scenarios of galaxy 
formation \citep[][]{2006MNRAS.366..499D,2007MNRAS.374..809D,2007MNRAS.375....2D}. But it 
also agrees with the recent results by \citet{2014ApJ...792...95C}, who report that the light-weighted ages
 of massive galaxies at $z<0.7$ indicate that they are noticeably younger than the age of the universe at 
all redshifts up to $z\sim 0.7$, which would suggest an effective single-burst star-formation epoch of $z\lesssim 1.5$ for them. 

\subsubsection{\bf GNCs by redshift bin: evidence of the progressive assembly of massive E--S0's at $0.8<z<1.5$}
\label{sec:gncsbyredshift}

\begin{figure*}
\begin{center}
\includegraphics[width=\textwidth,angle=0, bb = 0 36 481 185, clip]{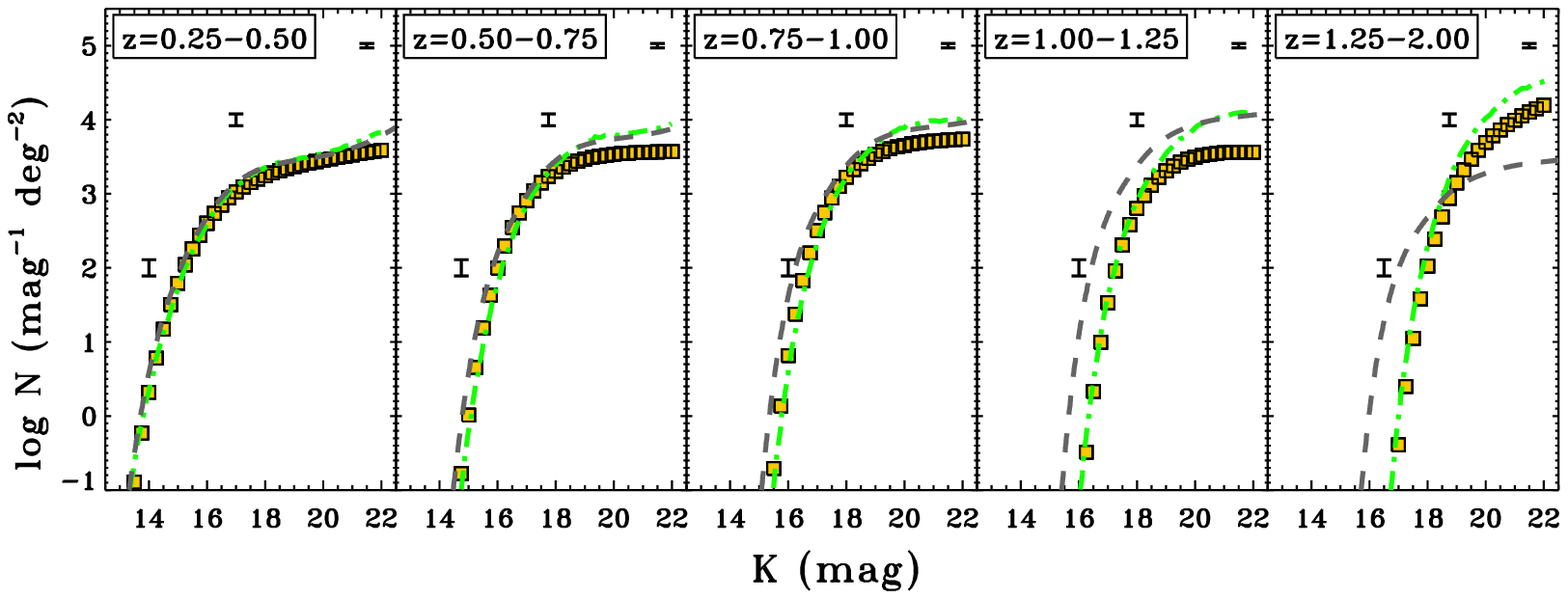}
\includegraphics[width=\textwidth,angle=0, bb = 0 0 481 223, clip]{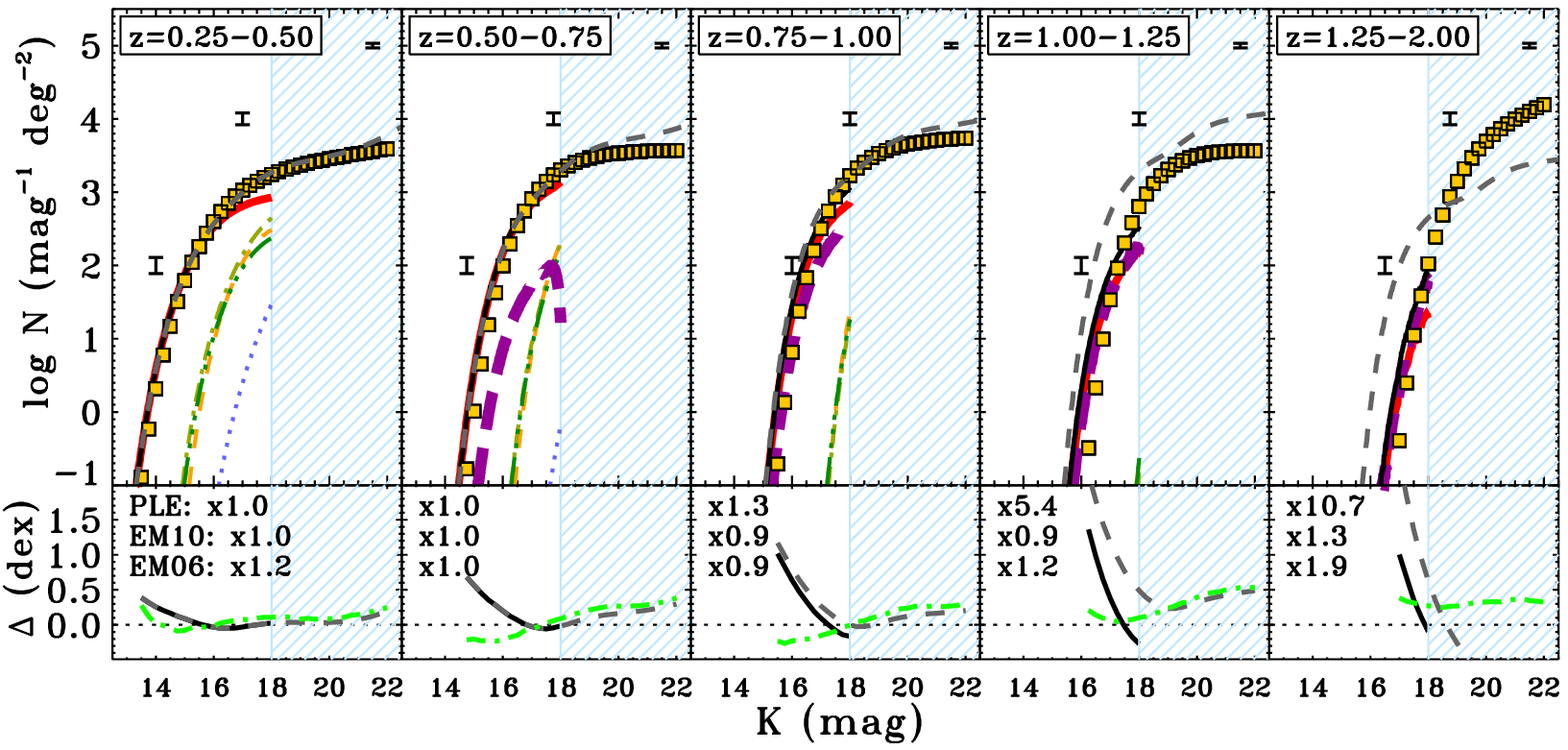}
\caption{Predictions of the models of the total $K$-band GNCs in different redshift bins up to $z=2$,
 compared to the real data from \citet{2009A&A...494...63B}.  \textbf{Top panels}: data compared to the 
PLE and EM06 models. \textbf{Middle panels}: data compared to the PLE and EM10 models. We also represent
 the contributions to the total GNCs in each redshift bin by morphological types expected by the EM10 model.
 The legend for these lines is the same as in the bottom panels of Fig.\,\ref{fig:ncstypesK2}. \textbf{Bottom
 panels}: differences in dex between the total GNCs in the $K$ band predicted by the PLE, EM06, and EM10 
models and the observational data in each redshift bin as a function of the apparent $K$-band magnitude. 
The factors written in each panel at the bottom represent the ratios of the total number of galaxies 
predicted by each model (PLE, EM10, and EM06) with respect to the total number of galaxies according to
 the observational data, all integrated down to the limiting magnitude set by the EM10 model at each 
redshift bin. \emph{Yellow squares}: observational data obtained by \citet{2009A&A...494...63B}. The 
typical observational errors at three magnitudes spanning the observational range have been 
overplotted, but displaced vertically from their data points for clarity. \emph{Grey dashed
 lines}: PLE model. \emph{Green dotted-dashed lines}: EM06 model. \emph{Black solid lines}: EM10 model.
 \emph{Region shaded in light blue}: Magnitudes for which the predictions of the EM10 model are not valid
 for the total GNCs in the $K$ band.  [\emph{A colour version is available in the online version.}] } \label{fig:ncstypesK}
\end{center}
\end{figure*}

Figure\,\ref{fig:ncstypesK} represents the predictions of the EM06 and EM10 models on the $K$-band GNCs 
by redshift bins, compared to the data derived by \citet{2009A&A...494...63B}. The predictions of the 
PLE model are shown in both cases for reference because the PLE scenario for the evolution of 
E--S0's since high redshifts is rejected when attending only to the total GNCs from NUV to NIR bands 
(\S\ref{sec:gncsmuultiwavelength}). In fact, the PLE model adequately fits real data up to $z\sim0.75$ in the 
figure, but it lies above the  observations at $z>0.75$, and predicts too many luminous galaxies and too few 
faint ones  at $1.25<z<2$. This model lacks a physical mechanism that converts massive galaxies into several low-mass components as we counter-evolve galaxy populations backwards in time.

The predictions of the EM06 model (top panels of the figure) are in better agreement with observations 
for $L>L^{\ast}$ galaxies than the PLE model, but they fail to reproduce the faint end of the GNC distribution 
from $0.5<z<0.75$ back to earlier epochs.  Therefore, assuming that most E--S0's formed at $z\sim 1.5$ provides a 
good approximation to the evolution of the bright end of the GNCs by redshift bin, but it does not work for the 
faint end. This suggests that the build-up of this galaxy population must have taken place at different epochs at
 $z\lesssim 1.5$, i.e.\ that the assembly of these galaxies at $z<1.5$ has been progressive. 

In the middle panels of Fig.\,\ref{fig:ncstypesK} we show that, within its limitations in magnitude 
range, the EM10 model reproduces the evolution of the massive end of the $K$-band GNCs by redshift bin. The 
growth of massive E--S0's is progressive in this model and basically takes place at $0.8<z<1.5$, whereas in the
 EM06 model (top panels) all these galaxies are inserted into the cosmic scenario at $z=1.5$. However, the
 progressive build-up of the massive E--S0's at $0.8<z<1.5$ in the EM10 model describes the bright end of the 
total GNCs by redshift bin in the $K$ band in a similar manner as the sudden build-up of these galaxies in the
 EM06 model. We explain the reason for this below.

We have also overplotted the contribution of the various galaxy types to the total GNCs expected by the EM10 model at each redshift bin in the middle panels of Fig.\,\ref{fig:ncstypesK}. The bright end of $K$-band GNCs is controlled by the E--S0's up to $z\sim 0.75$, but another population (those galaxies undergoing advanced stages of major mergers) starts to contribute similarly to the total GNCs from this epoch up to $z\sim 2$ in the model. Note that the addition of the GNCs of E--S0's and these advanced stages of major mergers at $0.75<z<2$ controls the bright end of the GNCs by redshift bin in the $K$ band. The difference from the  EM06 model in the top panels is that the EM06 model does not consider the existence of galaxies undergoing major mergers. However, the total $K$-band GNCs of E--S0's and galaxies in advanced stages of major mergers predicted by the EM10 almost coincides by chance with those of E--S0's expected by the EM06 model at all redshift bins, and they both reproduce the observed data similarly. Therefore, the EM06 model attributes the evolution observed in the bright end of the $K$-band GNCs by redshift bin as a mere colour evolution of the E--S0's up to $z\sim 1.5$ and to their vanishing at earlier epochs, whereas the EM10 model explains it through a progressive build-up of the massive E--S0's at $0.8<z<1.5$ through the major mergers reported strictly by observation. 

In the bottom panels of Fig.\,\ref{fig:ncstypesK}, we show the difference between the predictions of 
the three models and the data by \citet{2009A&A...494...63B} in dex as a function of the $K$ magnitude. The 
numbers in each panel correspond to the ratios of the total number of galaxies predicted by each model with 
respect to the total number of galaxies according to the observational data,  integrated down to the limiting
 magnitude set by the EM10 model in each redshift bin. Attending to the distribution of differences in dex, 
the predictions of the EM06 seem to fit the data better than the EM10 and PLE models in the range of validity 
of the EM10 model, but when the integrated ratios are compared, the EM10 model provides better predictions 
globally than the EM06 model at $z>1$. The EM06 model also predicts $K$-band GNCs higher than those expected
by the PLE model by $\sim 0.3$--0.5\,dex at faint magnitudes for $z>1$. 

The results of the EM10 model agree pretty well with observational evidence that massive galaxies 
have experienced passive evolution only since $z\sim 0.7$ according to their abundances of metals 
\citep{2014ApJ...792...95C}. These authors claim that newly quenched galaxies are added at $z<0.7$ only at 
the lower masses ($\log(M_\ast/\Msun)<10.5$), which means that the bulk of the build-up of the more massive ones 
must have taken place at $z>0.7$ (in agreement with the predictions of the EM10 model). This growth of the 
quiescent population only at the lowest masses is also supported by other studies \citep{2014arXiv1412.7162H}. However, other investigations find evidence of that the brightest quiescent members continue to accrete mass through dry major mergers in clusters at $z<0.6$ \citep{2015MNRAS.447.1491L}, so the real scenario of the build-up of these galaxies must be more complex.

In summary, $K$-band GNCs by redshift bin are an observable that directly discards the PLE model at $z>0.8$ 
for massive E--S0 galaxies and the `sudden' appearance of the whole population at $z\sim 1.5$, as proposed by the EM06 model (at least at the faint end). Instead, these GNC data imply a progressive build-up of  massive galaxies at $0.8<z<1.5$.

\subsubsection{\bf GNCs of E--S0's by redshift bin: this assembly involved $\sim 50$\% of present E--S0's at least}
\label{sec:gncsES0zbins}

\begin{figure*}
\begin{center}
\includegraphics*[width=\textwidth,angle=0]{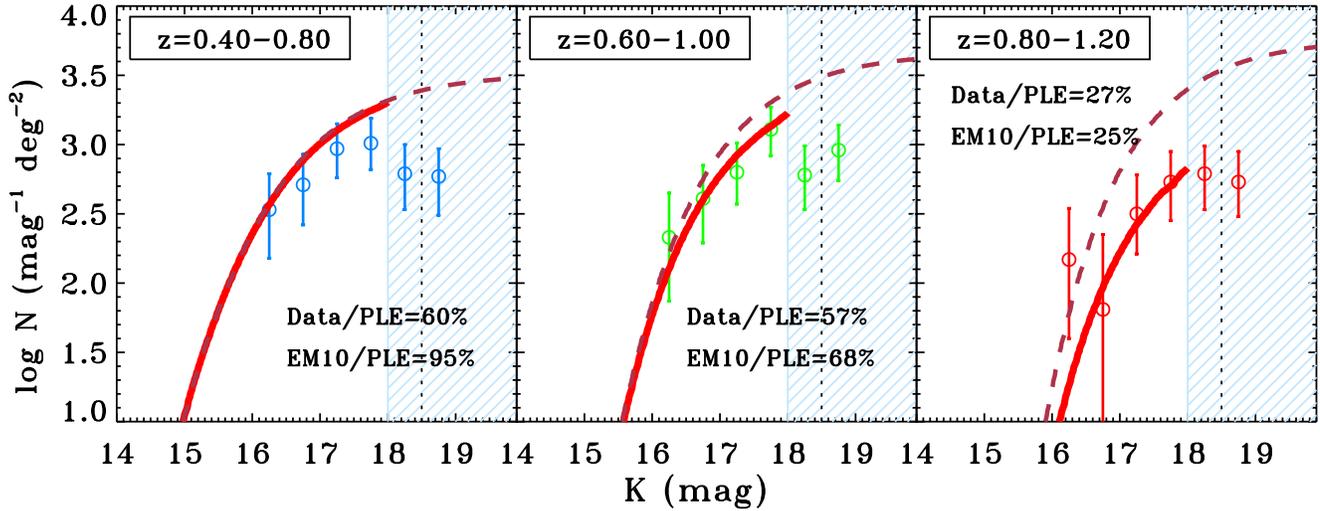}
\caption{Predictions of the EM10 model for the GNCs of E--S0's in wide, overlapping redshift bins in 
the $K$ band, compared to the data derived in the present study (see Fig.\,\ref{Fig:cuentasEs}). 
\emph{Circles}: GNCs of E--S0's for the redshift bin indicated in each panel. \emph{Red solid line}: 
GNCs of E--S0's predicted by the EM10 model in the corresponding redshift bin. \emph{Dark red dashed 
lines}:  GNCs of E--S0's predicted by the PLE model in each redshift bin. \emph{Blue shaded region}: 
Region where the predictions of the EM10 model are not valid for total $K$-band GNCs. \emph{Vertical 
dotted line}: limiting magnitude for 100\% completeness of our data. The percentages written 
in each panel represent the fraction of the total numbers of E--S0's derived from the observational 
data and expected by the EM10 model with respect to the number in the case of PLE, integrated from the 
lowest magnitude sampled by the observational data down to the limiting magnitude set by the EM10 model 
at each redshift bin.  [\emph{A colour version is available in the online version.}] \label{fig:mergers}}
\end{center}
\end{figure*}

In Fig.\,\ref{fig:mergers}, we compare the GNCs of E--S0's by wide redshift bins obtained in 
\S\ref{sec:countsES0} with the predictions of the EM10 model of the GNCs of E--S0's. The 
expectations for PLE are shown for reference. We also show the percentages represented by the total 
number of galaxies predicted by the EM10 model and the data with respect to the expectations of the 
PLE model for the magnitude range in which the EM10 model is valid. This figure shows that the EM10
 model provides predictions  much more in agreement with the data than PLE. The GNCs of E--S0's predicted 
by the EM10 model at the bright end are a factor of $\sim$$2$--3 times lower than the prediction of the PLE model at $0.8<z<1.2$. Comparison of the real data with the PLE model implies that the number density of massive E--S0's is lower than their local value by a factor of $\sim$$2$ at $0.6<z<1$ and $\sim$$4$ at $0.8<z<1.2$. Similar values are predicted by the EM10 model. The GNCs of E--S0's by redshift bin reveal that more than $\sim$$50$\% of present-day massive E--S0's have been progressively built up
 at $0.8<z<1.5$. The EM10 model attributes most of this assembly to the major mergers
 reported by observations (mainly through wet ones at high redshifts, see EM10).

The $K$-band GNCs of E--S0's by redshift bin thus support the key prediction of the EM10 model that the population of massive E--S0's decreases down to $\sim$$40$--50\% at $0.6<z<1$ with respect to the PLE extrapolation of their local population and down to $\sim 20$$--30\%$ at $0.8<z<1.2$. This numerical decrease of massive E--S0's agrees well with the estimates by \citet{2009ApJ...696.1554C} and 
\citet{2013A&A...558A..61M}. This finding implies that we may consider the bulk of the present-day massive E--S0 population to have been definitively assembled at $z\sim 0.8$--1, so that they may be considered to have been in place since $z\sim 0.6$ (i.e.\ only during the last $\sim 6$\,Gyr of cosmic history), in good agreement with the results by \citet{2014ApJ...792...95C} noted above. 

The EM10 model attributes this build-up entirely to major mergers. If this hypothesis turned out to be 
approximately true, it would mean that massive galaxies have experienced a growth in stellar mass of a factor 
$\times 2$ since $z\sim$$1$--1.2, which is in good agreement with the recent estimates by 
\citet{2015arXiv150102800S}. These authors also conclude that hierarchical models could explain this 
growth through major mergers, and that the rate expected for them is very similar to observational estimates, 
thereby supporting the scenario proposed by the EM10 model as well.

In summary, the $K$-band GNCs of E--S0's by redshift bin support the claim that present-day massive E--S0's 
experienced a noticeably progressive assembly at $0.8<z<1.5$, involving at least $\sim$$50$\% of their
 present-day number density. The EM10 model shows that it is feasible to explain this build-up just through 
the major mergers reported by observations up to $z\sim 1.5$.

\subsubsection{\bf GNCs by morphological type in different bands: major mergers as possible drivers
 of this assembly}
\label{sec:gncsmorphinbands}

\begin{figure*}
\begin{center}
\includegraphics*[width=\textwidth,angle=0]{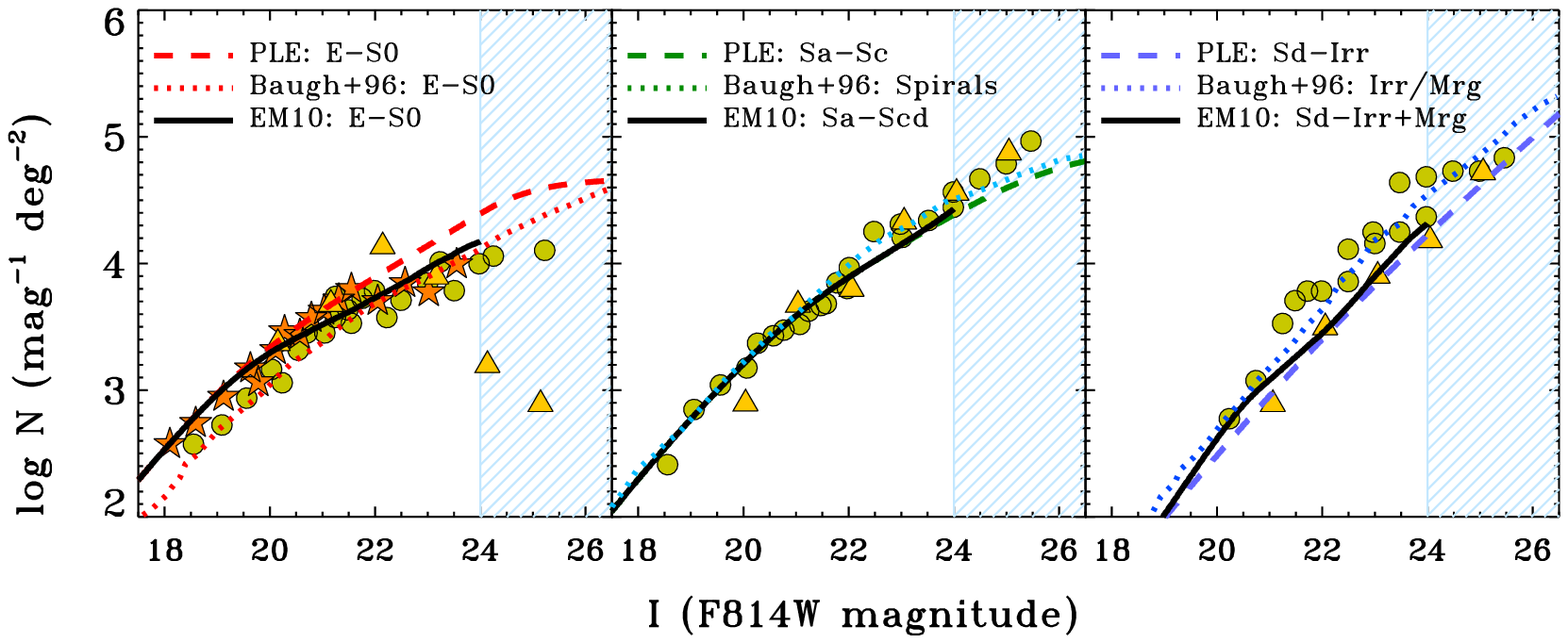}
\caption{Predictions of the PLE and EM10 models for the GNCs by morphological type in the $I$ band, 
compared to the data obtained by \citet[\emph{green circles}]{1998ApJ...496L..93D},
 \citet[\emph{orange stars}]{2000MNRAS.319..807P}, and \citet[\emph{yellow triangles}]{2000A&A...362..487V}. 
\emph{Left panel}: for the E--S0's. \emph{Middle panel}: for the Sa--Sc's. \emph{Right panel}: for the 
Sd--Irr's and mergers. \emph{Solid lines}: predictions of EM10 model for the galaxy type corresponding 
to each panel. \emph{Dashed lines}: predictions of the PLE model corresponding to each type, shown for 
reference. \emph{Dotted lines}: predictions of the semi-analytical model by \citet{1996MNRAS.282L..27B}
 for each galaxy type.  [\emph{A colour version is available in the online version.}]
 } \label{fig:ncsmorph}
\end{center}
\end{figure*}

In Fig.\,\ref{fig:ncsmorph}, we show the GNCs in the $I$ band by morphological type obtained by various
authors, compared to the predictions of several models. We have overplotted the expectations of the PLE and
 EM10 models, as well as the GNC predictions by galaxy type derived by \citet{1996MNRAS.282L..27B}
 from their semianalytic model \citep{1996MNRAS.283.1361B}. This model predicts the galaxy evolution 
starting from the power spectrum of primordial density fluctuations, and implementing the effects of 
different physical processes ruling galaxy formation and evolution, such as gravitational collapse, gas
 cooling, star formation, feedback, and galaxy mergers. This model adequately  reproduces the slope of the
 total GNCs in the $I$ band, but not in the $B$ band \citep[see Fig.\,1 in][]{1996MNRAS.282L..27B}. 

Figure\,\ref{fig:ncsmorph} demonstrates that GNCs by morphological type in optical bands are not very 
sensitive to galaxy evolution. Independently of the data dispersion in the figure, the three overplotted models
(which are very different among themselves) provide similar predictions for the GNCs of 
E--S0's, spirals, and Irr$+$Merger galaxies in the $I$ band. Therefore, little information can be 
obtained from morphological GNCs in optical bands.

\begin{figure*}
\begin{center}
\includegraphics[width=0.49\textwidth,angle=0, bb = 10 10 335 340, clip]{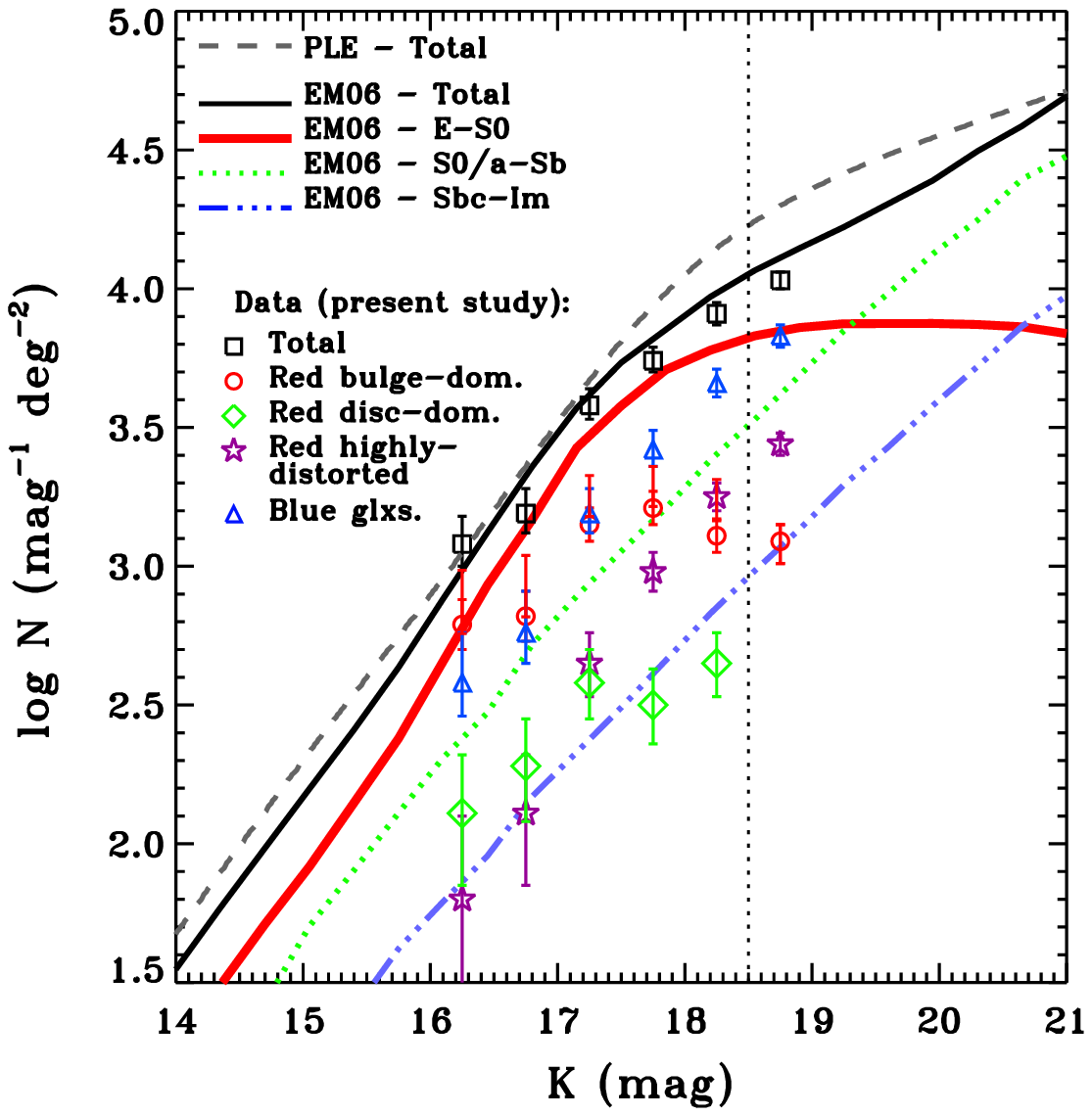}
\includegraphics[width=0.49\textwidth,angle=0, bb = 10 10 335 340, clip]{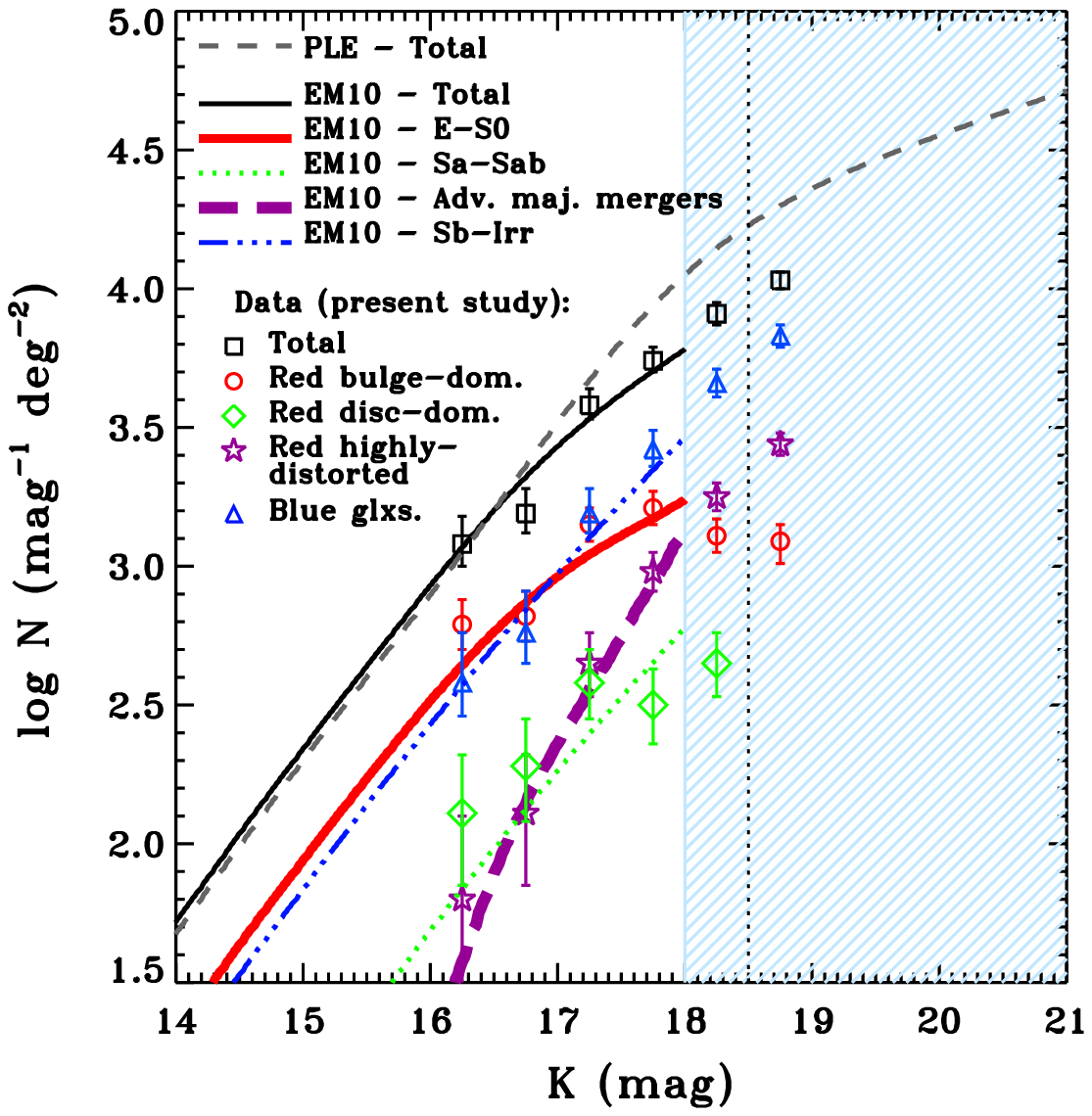}
\caption{$K$-band GNCs by morphological type derived in the present study (see Fig.\,\ref{Fig:cuentas}) 
compared to the predictions of the EM06 and EM10 models for analogous types (left and right panels, respectively). 
The prediction of the PLE model for total GNCs is shown in both panels for reference (\emph{grey dashed lines}).
 \emph{Vertical dotted line}: limiting magnitude for 100\% completeness of our data. \emph{Region shaded in
 blue in the right panel}: region where the predictions of the EM10 model are not valid for total $K$-band 
GNCs. Consult the legend in each panel.  [\emph{A colour version is available in the online version.}]
} \label{fig:comparacion}
\end{center}
\end{figure*}

However, we have found that morphological GNCs in the NIR bands are much more sensitive to the galaxy evolution 
experienced by each galaxy type than those in the optical bands. Figure\,\ref{fig:comparacion} shows the total 
GNCs by morphological types in the $K$ band derived in the present study (\S\ref{sec:counts}) compared 
to the predictions of the EM06 model (left panel) and the EM10 model (right panel). We have had to 
establish an ad hoc analogy between the data types defined in P13 (see \S\ref{sec:data}) and those defined 
in each model. The broad galaxy classes assumed by the EM06 model makes it difficult to compare the predictions 
of this model with the available data. We have identified the S0/a--Sb type in the model 
with red discs, but it is obvious that this class contains galaxies that are not strictly `red' (the Sb's), which should be included in the comparison with real blue galaxies instead. This explains why the model overpredicts the GNCs of red discs and underestimates the GNCs of blue galaxies at all magnitudes.  

We have shown that the simple scenario of the EM06 model can be rejected simply by attending to its 
predictions at the faint end of the total $K$-band GNCs by redshift bin (see Fig.\,\ref{fig:ncstypesK}). 
The disagreement between the data and the model predictions was interpreted as a sign that the bulk of 
low-mass ($L<L^\ast$) E--S0's cannot have formed at a short time period around $z\sim$$1.5$. Their definitive 
assembly was probably more gradual over the last $\sim$$9$\,Gyr. Attending now to the
 predictions of this model for the GNCs of E--S0's in Fig.\,\ref{fig:comparacion}, we can also refute a
 sudden formation of the bulk of bright E--S0's ($L>L^\ast$) at $z\sim 1.5$. The model clearly overpredicts
 their contribution at $K>16.5$\,mag. Moreover, the EM06 model does not provide any explanation for the 
existence of red highly distorted galaxies found in real data. It postulates the existence of initial stages 
of extreme dust extinction in the E--S0's after their appearance on the cosmic scene. But these galaxies remain E--S0's at those stages, they do not necessarily exhibit distorted morphology.

Nevertheless, the GNCs by morphological type predicted by the EM10 model in the $K$ band (right panel of
 Fig.\,\ref{fig:comparacion}) can approximately reproduce the observed distributions of all galaxy types 
identified in this study up to the limiting magnitude of validity of the model. To compare data and
 model predictions, we have identified the observed red regular spheroid-dominated galaxies with the E--S0's in
 the model, real blue galaxies with modelled Sb--Irr's, red regular disc-dominated galaxies with Sa--Sab's,
 and finally red highly distorted galaxies with the advanced stages of major mergers predicted by the model. 
The EM10 model is capable of explaining the shape and numbers of the GNCs of massive E--S0's up to 
$K\sim 18$\,mag, as well as the rise in GNCs of red distorted objects at faint magnitudes. Contrary to the 
EM06 model, the EM10 model naturally predicts high levels of dust extinction in these early phases of the 
assembled E--S0's because they are assumed to derive mostly from gas-rich major mergers, which are known
 to undergo intense starbursts at advanced stages of the interaction (see references in EM10). The success
 of the EM10 model is thus two-fold: first, it predicts the GNCs of bright E--S0's observed since 
$z\sim 1.5$, and, secondly, it offers an explanation of the appearance of a population of highly
 distorted galaxies on the red sequence at $z>0.8$ (marginal at lower redshifts), as the advanced stages of 
the major mergers reported by observations up to $z\sim 1.5$. This shows that it is feasible to explain the 
assembly of massive E--S0's at $0.8<z<1.5$ via major mergers. Note that it can also nearly explain the GNCs 
of red disc-dominated galaxies if they are identified with Sa--Sab's. 

The EM10 scenario is consistent with the results derived from the observed GNCs of Extremely Red Objects (EROs). 
EROs are considered as the counterparts of local $L> 2$--$3\times L^\ast$ galaxies at $0.7<z<1.5$, with the bulk
 of population at $z\sim 1$ \citep{2004A&A...421..821V}. In the framework of the model, the $L>L^\ast$ E--S0's
 derive from blue starforming galaxies through the major mergers that are observed to occur at $0.8<z<1.5$, 
meaning that these galaxies are undergoing advanced stages of these mergers at $0.6<z<1.2$, probably suffering 
intense dust extinction that reddens them significantly. Moreover, many other studies support that there must
 have been a significant assembly of red galaxies at the expense of the blue ones during the last 10\,Gyr 
\citep[see, e.g.,][P13]{2007ApJ...665..265F,2008ApJ...682..896K}. In particular, the model predictions agree 
with \citet{2013MNRAS.428.1460B}, who found that the early-type galaxies have been the predominant morphological 
class for $\log(M_\ast/\Msun) > 11$ since only $z\sim 1$.

\begin{figure*}
\begin{center}
\includegraphics[width=\textwidth,angle=0, bb = -5 0 481 160, clip]{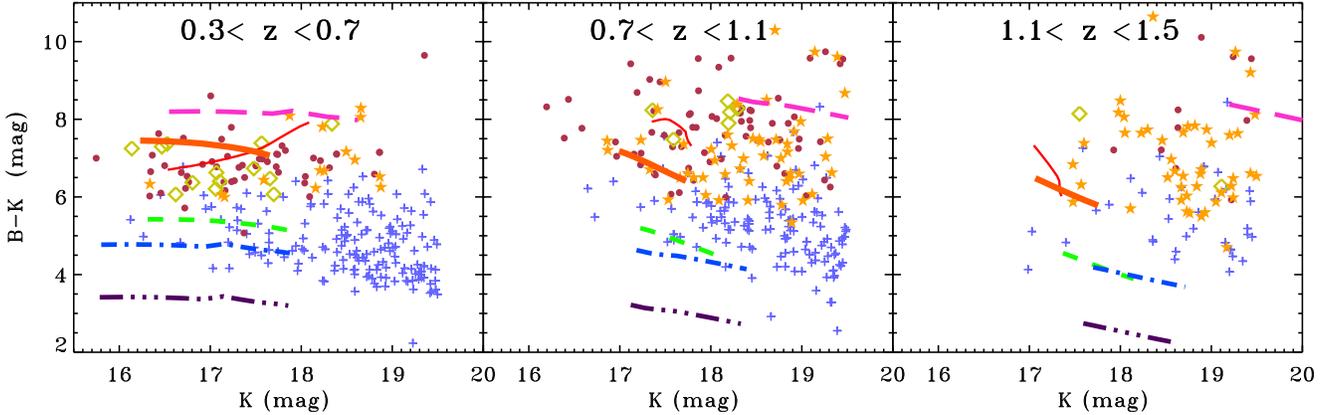}
\caption{Apparent colour--magnitude diagrams ($B-K$ vs.\,$K$) for our observational sample in the three wide redshift bins considered in P13 ($0.3<z<0.7$, $0.7<z<1.1$, and $1.1<z<1.5$). \emph{Red circles}: red regular spheroid-dominated galaxies (E--S0's). \emph{Green diamonds}: red regular disc-dominated galaxies. \emph{Yellow stars}: red highly distorted galaxies.  \emph{Blue crosses}: blue galaxies. \textbf{Lines}: theoretical trends followed by various galaxy types within the redshift interval of each frame, modelled assuming standard SFHs starting at $z_f=3$ and characteristic physical properties according to observations. They have been obtained with the stellar population synthesis models by \citet[see the text for details]{2003MNRAS.344.1000B}. Stellar masses of $M_* = 10^{11}$, $2\times 10^{11}$, and  $4\times10^{11}\Msun$ have been assumed for all models at $0.3<z<0.7$, $0.7<z<1.1$, and $1.1<z<1.5$ respectively. \emph{Red thin solid}: E--S0 galaxy. \emph{Orange thick solid}: Sa-Sab. \emph{Green dashed}: Sb-Sbc. \emph{Blue dot-dashed}: Sc-Scd. \emph{Purple three dots-dashed}: Sd-Irr. \emph{Pink long dashed}: Dust-reddened starburst galaxy with $A_V=3.0$\,mag.  [\emph{A colour version is available in the online version.}]
\/} \label{fig:cmdtypes}
\end{center}
\end{figure*}

The rise of spheroid-dominated galaxies at the expense of these highly distorted objects at the massive end of
 the red sequence at $0.8<z<1.5$ can be observed in the evolution of the relative fraction of both types in
 colour--magnitude diagrams. In Fig.\,\ref{fig:cmdtypes} we plot the apparent $B-K$ vs.\,$K$ diagrams for our
 blue and red samples in the three wide redshift bins defined in P13. The various morphological types of red
 galaxies are identified by distinct symbols. The expected traces for different galaxy types within each redshift
 interval have been overplotted. They correspond to the same models plotted in Fig.\,\ref{fig:cmd}, but
 different stellar masses have been assumed to scale the models in each redshift bin in order to overlap them
 with the data: $M_\ast= 1\times$, $2\times$, and $4\times$\,$10^{11}\Msun$ for $0.3<z<0.7$, $0.7<z<1.1$, and 
$1.1<z<1.5$ respectively. As times goes by, the relative presence of distorted objects in the red sequence
 decreases, whereas the relative fraction of regular spheroid- and disc-dominated galaxies rises. This figure also explains qualitatively why the appearance of the slope change in GNCs at $K\sim 17.5$\,mag. 
The GNCs at $K< 17$\,mag are numerically controlled mostly by the red objects at $0.3<z<1.1$ (compare the relative
 numbers of red and blue objects at $K< 17$\,mag in the three redshift bins). However, the number of blue 
objects at $0.3<z<1.1$ tends clearly  to dominate them at $K>18.5$. 

The EM10 model shows that it is feasible to attribute the build-up of massive E--S0's entirely to major
 mergers, but this scenario is an obvious oversimplification of reality. It does not account for additional 
evolutionary mechanisms, such as gas stripping and strangulation in clusters,
simple fading, or harassment \citep[see][for a review on the possible processes]{2012AdAst2012E..28A,2012ApJS..198....2K},
 that are known to have played a relevant role in the migration of galaxies from the blue 
cloud to the red sequence (mostly producing S0's). 
However, the evidence of the protagonism of major and minor merging in the formation of massive lenticulars has been growing in recent years, so the model could be providing a rough (but almost realistic) prediction of the
 global effects of mergers in this galaxy population  \citep[see, e.g.,][]{2006A&A...457...91E,2011MNRAS.412L...6B,2011MNRAS.412..684B,
2011A&A...533A.104E,2012A&A...547A..48E,2013A&A...552A..67E,2009A&A...496...51P,
2009A&A...501..437Y,2009A&A...507.1313H,2012MPLA...2730034H,2009A&A...496..381H,
2011MNRAS.412..684B,2011MNRAS.412L...6B,2014A&A...570A.103B,2014A&A...565A..31T,
2014ApJ...792...95C,2015A&A...573A..78Q}. The scenario proposed in the EM10 model
 is also supported by the recent results by \citet{2015ApJ...800..107W}, who indicate that
 the number evolution of the majority of massive early-type galaxies at $z>1.2$ seems to be
 driven by recent gas-rich mergers \citep[see also][]{2012ApJ...745..179W}.

In summary, GNCs by morphological type are much more sensitive to the evolution of massive E--S0's when derived in NIR bands than in the optical bands. The morphological GNCs of E--S0's and highly distorted objects on the red sequence are compatible with a scenario in which the build-up of massive E--S0's at $0.8<z<1.5$ has occurred through the major mergers reported by observations up
 to $z\sim 1.5$ (in good agreement with hierarchical theories of galaxy formation). The GNCs of E--S0's 
(total and by redshift bins) derived in the present study indicate that the slope change in the total 
GNCs at $K\sim 17.5$\,mag is nothing other than a vestige of the assembly of more than a half of
 present-day massive E--S0's at $0.8<z<1.5$, possibly driven by major mergers in their majority.

\section{Summary and Conclusions}
\label{sec:conclusions}

We have studied the galaxy population and evolutionary processes responsible for creating the slope
 change registered at $K\sim 17.5$\,mag in NIR GNCs. We have derived the contribution of various galaxy 
types to the total $K$-band GNCs at $0.3<z<1.5$, as well as the GNCs of E--S0's by redshift bin, in order 
to identify the populations that produce this feature. We have used the data and galaxy samples selected in
 P13. The GNCs and other data obtained from the literature have been compared with the expectations of 
a number of galaxy evolutionary models.

We show observationally  that the slope break in the $K$-band total GNCs is caused by a sharp change at $z<1.5$ in the galaxy population that controls them in terms of their numbers (from the quiescent E--S0's at $K< 17.5$\,mag to the blue star-forming discs at fainter magnitudes). The change is generated by a substantial flattening of the contribution of E--S0's to the total $K$-band GNCs at $z>0.6$, as compared to their rising trend at the bright end of the GNCs distribution.

From the comparison of the GNC data obtained here and from the literature with the models analysed, we have found the following results:

\begin{enumerate}

 \item We corroborate that models in which massive E--S0's evolve passively since high redshifts ($z>2$) cannot 
predict the slope change in the $K$ band GNCs and are discarded by attending only to the total GNCs from the NUV to
the NIR bands. On the contrary, we show that the models (such as EM06 and EM10) that assume a significant assembly of this galaxy population at $z< 1.5$ can simultaneously reproduce the total GNCs from the NUV to the NIR bands, 
including the slope break at $K\sim 17.5$\,mag. 

\item The total $K$-band GNCs by redshift bin reveal that massive galaxies (mostly E--S0's) have experienced 
a significant progressive assembly at $0.8<z<1.5$. 

\item The GNCs of E--S0's by redshift bin demonstrate that this assembly at $0.8<z<1.5$ involved at least  $\sim$$50$\% of the present-day massive E--S0 population ($L>L^\ast$). 

\item GNCs by morphological type are more sensitive to galaxy evolution when derived in the NIR than in the  red optical bands. 

\item The EM10 model, which considers massive E--S0's to have derived from the major mergers registered in observations, is capable of simultaneously reproducing the total GNCs from the NUV to the NIR bands, as well as total GNCs of E--S0's by redshift bin in the $K$ band. It can also reproduce the GNCs by morphological type if the following identification between the modelled and real types (as defined in P13) is established: the modelled E--S0's correspond to the observed red regular spheroid-dominated galaxies, Sa--Sab's 
in the model are the regular discs detected on the red sequence, Sb--Irr's are the blue galaxies in the data sample, 
and the advanced stages of major mergers predicted by the model are compared to the galaxies with distorted morphologies
 identified on the red sequence.  This shows that it is feasible to explain this assembly of massive E--S0's at
 $0.8<z<1.5$ (revealed by GNC data) only through major mergers.
\end{enumerate}

We conclude that the slope change in total NIR GNCs is thus a relic of the assembly of a substantial fraction of present-day massive E--S0's at $0.8<z<1.5$, in agreement with hierarchical theories of galaxy formation.

\section*{Acknowledgments}
The authors thank the anonymous referee for useful suggestions that have clearly improved the paper. 
We wish to acknowledge J.\,P.Gardner, and J.\,A.\,Newman and B.\,P.\,Moster for making their codes publicly
 available ({\tt NCMOD\/} and {\tt QUICKCV}, respectively). We thank N.\,Metcalfe for his compilation of GNC
 data, A.\,Grazian for providing us with his GNCs data in tabulated format, and Y.\,Matsuoka for interesting
 discussions on the topic of the paper. This work is based in part on services provided by the GAVO Data Center
 (http://dc.zah.uni-heidelberg.de). This research has made use of the NASA's Astrophysics Data System. 
This research has made use of the VizieR catalogue access tool, CDS, Strasbourg, France.
\newline
\newline
Supported by the Spanish Ministry of Economy and Competitiveness (MINECO) under projects AYA2006-12955,
 AYA2009-10368, AYA2012-30717, AYA2012-31277, AYA2013-48226, and AYA2013-43188, and by the Madrid 
Regional Government through the AstroMadrid Project (CAM S2009/ESP-1496, 
http://www.laeff.cab.inta-csic.es/\-projects/\-astromadrid/\-main/\-index.php). Funded by the Spanish MICINN under 
the Consolider-Ingenio 2010 Program grant CSD2006-0070: "First Science with the GTC" 
(http://www.iac.es/consolider-ingenio-gtc/), and by the Spanish programme of International
 Campus of Excellence Moncloa (CEI). 

\bibliography{elic0709_def.bib}
\bibliographystyle{mn2e}

\label{lastpage}
\end{document}